\documentclass[12pt]{article}
\begin{document}
\def\be{\begin{equation}}
\def\ee{\end{equation}}
\def\bea{\begin{eqnarray}}
\def\eea{\end{eqnarray}}
\def\nn{\nonumber}
\def\const{{\rm const}}
\def\v{\varphi}
\def\s{\sigma}
\def\vcl{\varphi_{\rm cl}}
\newcommand{\no}[1]{:\!#1\!:}
\def\la{\left\langle}
\def\ra{\right\rangle}
\def\d{\partial}
\def\se{S_{\rm eff}}
\def\tr{\rm Tr}
\hfill{NSF-ITP-01-27}

\hfill{UPRF-2001-08}
\begin{center}
{\large{\bf Matrix strings in a B-field}}

\bigskip
\bigskip
{\bf Gianluca Grignani \footnote{ Work supported in part by
INFN and MURST of Italy. Email: grignani$@$pg.infn.it} }

{\it Dipartimento di Fisica and Sezione I.N.F.N., Universit\`a di
Perugia,\\ Via A. Pascoli I-06123, Perugia, Italia}

\bigskip
{\bf Marta Orselli \footnote{ Work supported in part by
MURST of Italy. Email: orselli$@$fis.unipr.it} }

{\it Dipartimento di Fisica and I.N.F.N. Gruppo Collegato di Parma,\\ Parco
Area delle Scienze 7/A I-43100, Parma, Italia}

\bigskip

{\bf Gordon W. Semenoff \footnote{ Work
supported in part by NSERC of Canada and the National Science Foundation 
under Grant No. PHY99-07949. Permanent address: Department of 
Physics and Astronomy, University of British Columbia,
6224 Agricultural Road, Vancouver, British Columbia V6T 1Z1 Canada. Email: 
semenoff@physics.ubc.ca} }

{\it Institute for Theoretical Physics, University of California, 
Santa Barbara, California 93106-4030 USA}

\end{center}

\centerline{\bf Abstract}

We study the discrete light-cone quantization (DLCQ) of closed strings
in the background of Minkowski space-time and a constant Neveu-Schwarz
$B$-field.  For the Bosonic string, we identify the $B$-dependent part
of the thermodynamic free energy to all orders in string perturbation
theory.  For every genus, $B$ appears in a constraint in the path
integral which restricts the world-sheet geometries to those which are
branched covers of a certain torus.  This is the extension of a
previous result where the $B$-field was absent
~\cite{Grignani:2000zm}.  We then discuss the coupling of a $B$-field
to the Matrix model of $M$-theory. We show that, when we consider this
theory at finite temperature and in a finite $B$-field, the Matrix
variables are functions which live on a torus with the same
Teichm\"uller parameter as the one that we identified in string
theory. We show
explicitly that the thermodynamic partition function of the Matrix
string model in the limit of free strings reproduces the genus 1
thermodynamic partition function of type IIA string.  
This is strong evidence that the Matrix model can
reproduce perturbative string theory.  We
also find an interesting behavior of the Hagedorn temperature.

\newpage
\section{Introduction and Summary}

It is widely believed that the five different perturbatively
consistent superstring theories describe special points in the space
of vacua of a single dynamical structure called $M$-theory. However, 
the nature of the degrees of freedom and a detailed description of
the dynamics of $M$-theory are as yet unknown. What has
recently become apparent is that there are other points, besides the
perturbative string limits,  in the moduli
space of vacua where $M$-theory could be understood.  One of them is the
infinite momentum frame which is conjectured to be 
described by the Matrix model
~\cite{Banks:1997vh,Susskind:1997cw}.  Others are the various strong
field limits which produce field theories with non-commutative
geometry \cite{Seiberg:1999vs}, or non-commutative open string (NCOS)
theory \cite{Seiberg:2000ms},\cite{Klebanov:2000pp} or wrapped and 
non-relativistic closed string theory
models 
\cite{Danielsson:2000gi},
\cite{Gomis:2000bd},\cite{Danielsson:2001mu}. 
In this Paper we will compare two limits of
$M$-theory in an overlapping domain of validity - the Matrix model in
the limit where it should produce perturbative string theory and
perturbative string theory in the kinematical context, discrete light-cone
quantization (DLCQ),  which is
described by the Matrix model.  We will consider each model in a
background of a constant Neveu-Schwarz antisymmetric tensor field,
$B_{\mu\nu}$.  The coupling of this field to each model is simple and
its effect can be taken into account exactly.  This will give a
one-parameter comparison of the two theories and we will find
remarkable agreement.

The Matrix model \cite{Banks:1997vh,Susskind:1997cw} (for a recent
review see \cite{Taylor:2001vb}) with $N\times N$ matrices and gauge
group $SU(N)$ is conjectured to describe $M$-theory on a background of
eleven dimensional Minkowski space with a compact null direction,
$X^+\sim X^++2\pi R$ and $N$ units of null momentum, $P^-=N/R$.

Type IIA superstrings can be obtained from $M$-theory by compactifying a
space direction. This compactification adds a dimension to the matrix
model \cite{Taylor:1997ik}.  The resulting Matrix string theory
\cite{Motl:1997th},
\cite{Dijkgraaf:1997vv} is 1+1-dimensional maximally supersymmetric 
Yang-Mills theory.  With
the appropriate identification of degrees of freedom, it should be a
non-perturbative formulation of type IIA superstrings on a space with
a compact null direction.  Dijkgraaf, Verlinde and Verlinde 
\cite{Dijkgraaf:1997vv}  argued that perturbative string
theory is described by the moduli space of classical vacua of the
1+1-dimensional Yang-Mills theory.  In this moduli space, the degrees
of freedom are mutually commuting matrices.  The string degrees of
freedom are the simultaneous eigenvalues of the matrices.  The correct
description of the dynamics of these eigenvalues in the limit 
which produces free strings is a super-conformal
field theory on a symmetric orbifold $(R^8)^N/S_N$.  They also showed
that the first correction to the free string Hamiltonian is an
irrelevant operator which precisely reproduces the Mandelstam
three-string vertex.  Elaborations on this limit have been discussed
in detail in the literature
~\cite{Wynter:1997yb}-\cite{Kostov:1998pg}, \cite{Grignani:2000zm}.

In the following, we will provide further support for these ideas by
comparing that limit of matrix string theory which should produce
weakly coupled strings with discrete light-cone quantized (DLCQ)
perturbative string theory.  We will be particularly interested in
examining the effect of coupling a constant background $B_{\mu\nu}$ to
the Matrix string model and comparing the perturbative string limit to
the DLCQ type IIA superstring in a background $B$-field.  We will find
a remarkable agreement between the two.  This is an elaboration of our
previous results in ~\cite{Grignani:1999sp},\cite{Grignani:2000zm}
where a similar comparison was made in the absence of $B_{\mu\nu}$.
It provides further evidence in support of the Matrix model
conjecture.

We will be interested in the thermodynamic partition function of
matrix string theory.  To form the thermodynamic partition function,
we must identify the energy.  In the Matrix model, one of the
light-cone momenta is given by $P^-=N/R$ where $R$ is the
compactification radius of the light-cone and the other one is given
by $P^+=H$, the Hamiltonian of 1+1-dimensional supersymmetric
Yang-Mills theory with gauge group $U(N)$.  The energy is given by
$P^0=(P^++P^-)/\sqrt{2} =(N/R+H)/\sqrt{2}$ which can be used to form
the partition function
\begin{equation}
Z(\beta)= {\rm Tr}( e^{-\beta P^0})=
\sum_{N=0}^\infty e^{-\beta N/\sqrt{2}R}{\rm Tr}( e^{-\beta H/\sqrt{2} })
\label{partitionsum}
\end{equation}
where $\beta=1/T$ is the inverse temperature.  The sum over $N$ is the
trace over the spectrum of $P^-$.  The remaining trace is the
thermodynamic partition function of the 1+1-dimensional super
Yang-Mills theory with gauge group U(N) and inverse temperature
$\beta/\sqrt{2}$.  It is given by the usual finite temperature field
theory path integral with compact Euclidean time where the
supersymmetry of the model is broken by the anti-periodic boundary
conditions for fermions.
It will turn out that the matrix string model partition function
(\ref{partitionsum}) contains a functional integral over matrix-valued
fields which live on a particular 2-torus,
characterized by Teichm\"uller parameter 
\begin{equation}
\tau_0=i\frac{(
1-2\pi\alpha'iB^E)}{\nu}
\label{tauzero}
\end{equation}
where
\begin{equation}
\nu=\frac{4\pi\alpha'}{\sqrt{2}\beta R}
\label{nu}
\end{equation}
and $B^E$ is the Euclidean version of a component of the $B$-field.
We will find that this underlying torus also emerges in the string
theory in an interesting way.

We shall begin by studying the closed bosonic string and the IIA
superstring in a constant external $B_{\mu\nu}$.  In order to compare
with the Matrix model, we do discrete light-cone quantization (DLCQ)
of the string.  We shall consider the
thermodynamic partition function.  Studying either string theory or
Matrix theory at finite temperature compactifies Euclidean time,
$X^0\sim X^0+\beta$.    In the IIA
string theory, this gives a second compactification of the target
space: both the null Minkowski direction and Euclidean time are then
compact.  In \cite{Grignani:1999sp},\cite{Grignani:2000zm} it was seen
that this double compactification results in an interesting constraint
on geometries of world-sheets in the path integral formulation of the
string.  Here, we shall see how this constraint is modified by
the presence of a $B$-field.The result will be that the integral over
all geometries of Riemann surfaces in the string path integral is 
reduced to an integral over those Riemann surfaces which are branched
covers of a torus with Teichm\"uller parameter given by $\tau_0$ in 
(\ref{tauzero}).

Without compactifications of space-time, and in the absence of D-branes, 
closed strings would not couple to a constant $B$-field.
In fact, a constant $B$-field is gauge equivalent
to a constant electromagnetic field and the electric charges that would
couple to it live at the ends of open
strings.  Closed strings are neutral and are normally 
unaffected by an electromagnetic field.  
However, when some space-time directions are compact,
the states where the closed string wraps the compact dimension can couple to $B$.  
The coupling amounts to a constant shift of energies and momenta of the
wrapped strings.  Its origin can intuitively be understood by imagining a closed
string as being made from an open string by fusing its ends.  If the open string
carries charges of opposite sign on its endpoints, the process of creating a small
open string, wrapping it around the compact dimension, then fusing the ends together
obtains a contribution to its energy from transporting the charged
endpoints in a constant electric field.  The energy shift is 
$ 2\pi Rn\cdot B$ where $B$ is the component of electric 
field in the compact direction, $R$ is the compactification
radius and $n$ is the number of times the resulting closed string wraps the compact
direction.    A constant shift in energy is a chemical potential and 
an electric $B$-field
should therefore induce a finite density of wrapped closed strings.   There
would also be a similar shifts in momenta of wrapped states 
coming from the magnetic components of the $B$-field.If more directions were compact, the
presence of the $B$-field would lead to a higher dimensional non-commutative
Yang-Mills theory ~\cite{Connes:1998cr}, \cite{Douglas:1998fm}, \cite{Cheung:1998nr}.

Consider a closed bosonic string 
propagating on 26-dimensional Minkowski space with a constant background 
$B$-field.  The action is given by
$$
S=-\frac{1}{4\pi\alpha'}\int d^2\sigma\partial^a X^\mu\partial_a X_\mu
+\frac{1}{2}B_{\mu\nu}\int d^2\sigma \epsilon^{ab}\partial_a X^\mu \partial_b
X^\nu
$$
The $B$-field contributes a total derivative term to the action.  The equations
of motion do not depend on $B$ and have their usual form,
$$
\partial_+\partial_- X^\mu=0
$$
$$
\partial_+ X^\mu\partial_+ X_\mu = 0
~~,~~
\partial_- X^\mu \partial_- X_\mu = 0
$$
Noether currents can depend on $B$. For example,
total momentum which is the Noether charge for space-time
translation invariance, has the form
\begin{equation}
P^\mu =\frac{1}{2\pi\alpha'}\int_0^\pi d\sigma \partial_\tau X^\mu
+B^{\mu}_{~\nu}\int_0^\pi d\sigma\partial_\sigma X^\nu
\label{momenta}
\end{equation}
and we see that it is modified by $B$ only if for some $\mu$, $\int
d\sigma\partial_\sigma X^\mu \neq 0$.  This can happen when the string
wraps a compactified dimension.

Let us consider the simple case where one of the spatial 
dimensions is compact,
$X^{1}\sim X^{1}+2\pi R_{1}$ and where the only non-zero component of $B$ is 
$B_{01}\equiv B$.
The mass-shell and level matching conditions are
\begin{equation}
\left( P^0-2\pi BR_{1}n\right)^2-{\vec p}^{~2}= \left(\frac{m}{R_{1}}\right)^2
+\left( \frac{nR_{1}}{\alpha'}\right)^2+\frac{2}{\alpha'}\left( N+\tilde N-2
\right)
\label{massshell}
\end{equation}
\begin{equation}
N-\tilde N = mn
\label{levelmatching}
\end{equation}
Here $m/R_{1}$ is the quantized momentum in the compact dimension.
The integer $n$ is the number of times the string wraps the compact
dimension.  The shift of the energy by the $B$-dependent term is a
result of the shift of the momentum in (\ref{momenta}).  We see that,
for fixed $n$, the $B$-field affects the spectrum like a chemical
potential, $2\pi BR_{1}n$ in (\ref{massshell}).

The mass-shell condition (\ref{massshell}) is solved by light-cone momenta, 
$P^\pm=\frac{1}{\sqrt{2}}\left(P^0\pm P^1\right)$, 
\begin{eqnarray}
P^-=\frac{1}{\sqrt{2}}\left( 2\pi BnR_{1}+
\sqrt{ \left(\frac{m}{R_{1}}\right)^2
+\left( \frac{nR_{1}}{\alpha'}\right)^2+\vec p^{~2}
+\frac{2}{\alpha'}(N+\tilde N-2)}
+\frac{m}{R_{1}}\right)
\nonumber \\
P^+=\frac{1}{\sqrt{2}}\left( 2\pi BnR_{1}+
\sqrt{\left(\frac{m}{R_{1}}\right)^2
+\left( \frac{nR_{1}}{\alpha'}\right)^2 + 
\vec p^{~2}+\frac{2}{\alpha'}(N+\tilde N-2)}
-\frac{m}{R_{1}}\right)
\end{eqnarray}
We can obtain a compactified null direction from a compactified
spatial direction by an infinite boost along the compact direction
together with an infinite rescaling of $R_1$ 
~\cite{Seiberg:1997ad}. We consider the state where the 1-component of
the momentum is large and negative and boost in the positive 1
direction.  The light-cone momenta in the boosted frame are scaled by
the factors $$
\tilde P^-=\lim_{R_{1}\to 0} \frac{R_{1}}{\sqrt{2}R}P^-
~~,~~
\tilde P^+=\lim_{R_{1}\to 0} \frac{\sqrt{2}R}{R_{1}}P^+
$$
to get
\begin{equation}
\tilde P^-= \frac{m}{R}
~~,~~
\tilde P^+ = 2\pi B nR +\frac{R}{2m}\left(\vec p^{~2}+\frac{4}{\alpha'}
(N+\tilde N-2)\right)
\label{boostedmom}
\end{equation}
Now, $R$ is the light-cone compactification radius, $X^+\sim X^++2\pi
R$.  The $B$-field contributes a chemical potential-like shift to the
light-cone energy $P^+$.  The only other place that the wrapping
number, $n$, appears is in the level matching condition
(\ref{levelmatching}) which is unchanged.
Note that if we combine (\ref{boostedmom}) and (\ref{levelmatching}), the
mass operator has the form
\begin{equation}
2P^+P^--\vec P^{~2}=\frac{2}{\alpha'}\left( 1+2\pi\alpha'B\right)
\left(N-1\right)
+\frac{2}{\alpha'}\left( 1-2\pi\alpha'B\right)\left(\tilde N-1\right)
\label{tensionshift}
\end{equation}
In this formula, the string tensions of left and right-movers are
shifted by factors of $(1+2\pi\alpha' B)$ and $(1-2\pi\alpha'B)$,
respectively.

We are interested in seeing how this spectrum arises in the appropriate limit
of Matrix string theory.  We shall consider 
the case where $B_{01}=B_{-+}\equiv B$ is the only
non-zero component of the $B$-field and where it is a constant.  To find
the action of Matrix string theory in this $B$-field, our starting
point is the action for D1-branes on a space-time with the
$X^1$-direction compactified and with a background metric

\be G_{\mu\nu}= \left( \matrix{
-1+(2\pi\alpha'B)^2 & 2\pi\alpha'B & 0 & ... \cr 2\pi\alpha' B & 1 & 0
& ... \cr 0 & 0 & 1 & ... \cr ....  & ...  & ...& ...\cr } \right) 
\label{metric1}
\ee
and no $B$-field
$$
B_{\mu\nu}=0 
$$
The correct D1-brane action can be found by dimensionally reducing
10-dimensional supersymmetric Yang-Mills theory with the above
space-time metric to obtain maximally supersymmetric 1+1-dimensional
Yang-Mills theory whose 2-dimensional space-time metric is the upper
left-hand corner of $G_{\mu\nu}$.

Using the Buscher rules \cite{Buscher:1988qj,Buscher:1987sk} for
$T$-duality, this is equivalent to D0-branes on a space with the same
compactified direction with dual radius, the Minkowski metric
$G_{\mu\nu}={\rm diag}(-1,1,1,1,...)$ and non-zero $B_{01}=B$-field.

Then, a combination of arguments following Seiberg
\cite{Seiberg:1997ad} and Sen \cite{Sen:1998we} and Dijkgraaf,
Verlinde and Verlinde \cite{Dijkgraaf:1997vv} can be used to show
that, with the appropriate identification of parameters, this D0-brane
action is also the Matrix string model which should describe Matrix
strings in a $B$-field.  Note that, under the boost to the infinite
momentum frame which is required to make this identification, the
component of the $B$-field that we are interested in, $B_{01}=B_{+-}$,
is invariant.
 
The linearized coupling of an external $B$-field to both the Matrix
model and the Matrix string model has been found before
\cite{Schiappa:2000dv,Taylor:2000pr}.  We find that when $B_{\mu\nu}$
is a constant, the full $B$-dependence of the Matrix string model action
has linear and quadratic terms in $B$.  The linear term agrees with
the one found in \cite{Schiappa:2000dv}.  From the Matrix string
action, we can also deduce how the $B$-field appears in the Matrix model
action.  There is also a linear and quadratic term and the linear
term agrees with the coupling found in \cite{Taylor:2000pr}.

Weakly coupled 
string theory has a density of states which grows exponentially with 
energy ~\cite{Hagedorn:1965st}.  This
gives rise to a maximum temperature, called the Hagedorn temperature, 
beyond which
a gas of strings cannot be heated.  This Hagedorn temperature is also 
sometimes
interpreted as a phase transition 
\cite{Sathiapalan:1987db,Kogan:1987jd,Atick:1988si}.

It is interesting to ask the question where a Hagedorn-like behavior 
could arise
in (\ref{partitionsum}).  Since the last factor is just the partition sum
of a field theory, it is clear that the only possible source of divergence
which could give a Hagedorn temperature is in the summation over $N$. 
The convergence
of the sum over $N$ is governed by the large $N$  
limit of the gauge theory. Indeed,
there is a rough classification of the behaviors that can occur
according to the 
three possible asymptotic behaviors
for this sum, depending on the state of the
gauge theory in the large $N$ limit.  The free energy, $F$, 
of the gauge theory is 
defined by
$$
{\rm Tr}e^{-\beta H/\sqrt{2} }~=~ e^{-\beta F/\sqrt{2}}
$$
As a function of $N$ in the large $N$ limit, the free energy can generally
be expected to be negative and proportional to a power of $N$,
$$
\beta F \sim -N^a\cdot\ldots
$$
There is a simple classification of the behavior of the sum over $N$ which 
depends on the exponent, $a$, 
\begin{equation}
\matrix{ a<1 & {\rm confinement} \cr  
a=1,~ \beta F\sim-N\beta_H/R & {\rm free ~ strings} \cr
a>1 ~ {\it e.g.}~\beta F\sim -N^2\cdot\ldots & {\rm deconfined} \cr }
\label{cases}
\end{equation}

An example of the first case is the gauge theory in a confining phase.
The large $N$ limit of its free energy is of order one.  Then, the sum
over $N$ in (\ref{partitionsum}) will be convergent and the partition
function will be field theory-like with a field theoretical asymptotic
density of states.

In the second case in (\ref{cases}), which is the one that we will
study in this Paper, the eigenvalues of the matrices dominate and
there are $N$ of them, thus the free energy of order $N$.  This is the
limit of free long strings found by Dijkgraaf, Verlinde and Verlinde
\cite{Dijkgraaf:1997vv}.  We emphasize that, in this Paper, we will
not discuss how this limit is obtained or whether it exists as a
bona-fide phase of the 1+1-dimensional gauge theory.  We will assume
that it is obtained in the limit of strong gauge theory coupling and
examine the consequences.  In this case, the partition sum will
diverge when $\beta<\beta_H$ with $\beta_H=1/T_H$ defining the
Hagedorn temperature.

In the third case in (\ref{cases}), $\beta F\sim -N^2\ldots$, 
which is what is generically expected in this supersymmetric gauge
theory in the large $N$ limit.  If the theory is in a deconfined
or screening phase, the free energy is proportional to $-N^2$ in 
large $N$ 't Hooft
limit where $g_{\hbox{\tiny{\rm YM}}}^2N$ is held fixed.  
Then, the partition sum does not 
converge for any value of the parameters.  We regard this as a  
reflection of the Jeans instability of hot flat space.  States of the Matrix model
which have  free energy of order
$-N^2$ in the large $N$ limit correspond to string theory
black holes.  Divergence of the
partition sum is due to the nucleation of black holes which 
dominate the entropy at any temperature \cite{Rozali}.  An explicit
evaluation of the perturbative limit of the Yang-Mills theory can be
found in refs.\cite{Ambjorn:1999zt},\cite{Ambjorn:2000yr}.

\subsection{Results:}

First of all, to illustrate the point, we quantize the Bosonic closed
string using DLCQ and operator methods.  We find the spectrum and use
it to compute the thermodynamic partition function for free strings.
This generalizes previous work
\cite{Grignani:1999sp} to a constant $B$-field.  We find that, as in
\cite{Grignani:1999sp}, the free energy is given by a discrete sum
over Teichm\"uller parameters,
\begin{eqnarray}
\frac{F}{V}=-\sum_{n=1}^\infty \sum_{k=1}^\infty \sum_{s=0}^{k-1}
\frac{1}{k^2} \left(\frac{1}{4\pi^2\alpha'\hat\tau_2}\right)^{13}~
\exp\left(-\frac{n^2\beta^2}{4\pi\alpha'\hat\tau_2}\right)
\left(\eta\left(\tau_-\right)\right)^{-24}
\left(\bar{\eta}\left(\tau_+\right)\right)^{-24}
\label{fsfsfs}
\end{eqnarray}
where 
\begin{equation}
\tau_{+}= 
\frac{s}{k}+i\frac{(1 + 2\pi\alpha'B)}{\nu}\frac{ n}{k}
\label{tauplus}
\end{equation}
\begin{equation}
\tau_{-}= 
\frac{s}{k}+i\frac{(1 - 2\pi\alpha'B)}{\nu}\frac{ n}{k}
\label{tauminus}
\end{equation}
and 
$
\nu=\frac{4\pi\alpha'}{\sqrt{2}\beta R}
$, the parameter in (\ref{nu}).

The integers take values $k,n=1,2,3,...$ and $s=0,1,...,k-1$.
Also, 
\begin{equation}
\hat\tau_2= \frac{n}
{k\nu}
\label{tautwo}
\end{equation} 
is the average of the imaginary parts of 
$\tau_+$ and $\tau_-$.
In spite of the $B$-dependent difference between $\tau_+$ and
$\tau_-$, this partition function is real and exhibits the expected
modular invariance.  It has a symmetry under the replacement
$$
\sqrt{2}\pi \beta RB \to \sqrt{2}\pi \beta RB +2\pi i\cdot~{\rm integer}
$$
This symmetry is related to a large gauge invariance of the string theory on
a toroidal space, $B\rightarrow B+2\pi i \cdot(\rm integer)/V_T$, where 
$V_T$ is
the volume of the torus, is a gauge transformation.  The free energy exhibits
a symmetry under this gauge transformation.  

We then show how the same result can be obtained in a covariant
quantization using the path integral where the worldsheets have genus
1. In this case the simultaneous compactification of the Minkowski
space null direction and Euclidean time involves an periodic
identification of coordinates that contains complex numbers.  It
further requires that we use a Euclidean $B^E$-field which is related
to the physical Minkowski one by $B^E_{01}=-iB_{01}$.  We can
re-obtain the Minkowski one by analytic continuation only after the
computation of the path integral is done.  Our result then reproduces
the one that we obtained using operator methods quoted in
(\ref{fsfsfs}) above.

We then use the covariant path integral formulation to find the
$B$-dependence of the free energy of the Bosonic string at all genera.  The
result is that, at a given genus, $B^E_{01}$ appears only in a
constraint on the integration over world-sheets in the path integral.
At genus $g$, this constraint restricts the integral over world-sheets
to those whose period matrix obeys a condition (see Section 4 for
notation)
\begin{equation}
\sum_{i=1}^g\left(k_i+\tau_0 m_i\right)\Omega_{ij}-\left(
s_j+\tau_0n_j\right)=0
\label{constraint}
\end{equation}
where $m_i,n_i,k_i,s_i$ (i=1,...,g) are 4g integers.  This is the
condition that the Riemann surface with period matrix $\Omega_{ij}$ is
the branched cover of a torus with Teichm\"uller parameter
$$
\tau_0=i\frac{1-2\pi \alpha'iB^E}{\nu}
$$
identical to that in (\ref{tauzero}).
This result, with $B^E=0$ was found in \cite{Grignani:2000zm}.  The
constraint (\ref{constraint}) restricts the integration over all
Riemann surfaces in the path integral computation of the free energy
to an integral over those Riemann surfaces which are branched covers
of a particular torus.  The only place that the constant $B$-field
enters the partition function is in the geometry of the underlying
torus.

Note that the two different parameters $\tau_+$ and $\tau_-$ defined
in (\ref{tauplus}) and (\ref{tauminus}), respectively, have underlying
$\tau$-parameters which are obtained
by replacing $B^E$ by $-iB$ in $\tau_0$ and $\bar{\tau}_0$, respectively.

We then show that essentially the same constraint occurs for the
Green-Schwarz superstring in the same geometrical setting.  For the
superstring, we present an operator quantization (DLCQ) which gets the
genus 1 contribution to the partition function.  
The essential formula for the genus 1 contribution is 
$$ 
\frac{F}{V}=-\sum_{\stackrel {n=1} {n ~\rm odd} }^\infty 
\sum_{k=1}^\infty 
\sum_{s=0}^{k-1}\frac{1}{k^2}\left( \frac{1}{4\pi^2\alpha'\hat{\tau}_2} 
\right)^5 
2^9 \left(\theta_4\left(0,2\tau_- \right)\right)^{-8} 
\left(\bar{\theta}_4\left(0,2\tau_+  \right)\right)^{-8} 
e^{-n^2\beta^2/4\pi\alpha'\hat{\tau}_2} 
$$
An investigation of the superstring at 
higher genera is not given here.  Obtaining
the effect of compactification of the null direction on the finite
temperature partition function similar to (\ref{constraint}), but for
the super-geometry of the world-sheet should be
straightforward.

We then compare our results for the superstring with what we would
expect to obtain from the Matrix string model in the same $B$-field.
We show that, if we were to compute the finite temperature partition
function of the Matrix string model, we must do a path integral for
2-dimensional supersymmetric Yang-Mills theory where the Yang-Mills
field variables live on a torus with Teichm\"uller parameter
$\tau_0$. This parameter is identical to the $\tau_0$
which occurred in the string, given in equation (\ref{tauzero}).

The hypothesis is that the string degrees of freedom are the
simultaneous eigenvalues of the matrices.  If the matrix elements are
doubly periodic functions on the torus, the eigenvalues of the
matrices live on branched covers of the torus.  We have shown that it
is precisely these branched covers which we should expect to become
the world-sheets of strings in perturbation theory, where the genus of
the branched cover is the genus of the worldsheet.  Thus, we can add
the conjecture that summing the Matrix string model partition function
over the moduli space of branched covers with the appropriate measure
should produce string perturbation theory.  This has not yet been done
in detail beyond genus 1.  However, we can check the example of genus
1 explicitly and we find that the partition function of the Matrix
string model - where we sum over all genus 1 (unbranched covers) of
the basic torus - and the IIA Green-Schwarz superstring are indeed
identical.  This was done in the absence of $B$-field in
\cite{Grignani:1999sp}.  The extension in this
Paper to the case with a constant $B$-field is interesting because the
$B$-field modifies the geometry of both the Matrix model and the
quantized string in a simple way.  Seeing that this change maps
correctly through the limit of the Matrix model which produces
perturbative strings is a non-trivial check of the Matrix string model.

A by-product of our analysis is an expression for the Matrix model action
in a constant $B$-field in the $(0,1)$ direction,
\begin{eqnarray}
S=\int dt\left\{ \frac{1}{2g\sqrt{\alpha'}}{\rm Tr}
\left[\left(D_{\tau}X^a +B_{01}
[X^1,X^a]\right)^2-\frac{1}{(2\pi\alpha')^2}[X^a,X^b]^2\right]  
\right. \nonumber \\
\left.
-\frac{i}{2\pi}\left( \psi^TD_{\tau}\psi +2B_{01}\psi^T[X^1,\psi]
 -\frac{i}{2\pi\alpha'}\psi^T\gamma_a[X^a,\psi]\right)\right\}
\end{eqnarray}

Finally, we examine the Hagedorn phase transition in a finite $B$-field.
We observe the interesting fact that for the Discrete light-cone
quantized closed bosonic string, when one of the directions in which
there is electric $B$-field is compactified, the Hagedorn temperature
depends on $B$: 
$$ 
T_H= \sqrt{ \frac{
1-(2\pi\alpha'B)^2}{16\pi^2\alpha'}} $$ 
This is remarkable in that it doesn't depend on the compactification
radius, so it must hold even if the compactification radius is
arbitrarily large.  Of course, without the compactification in the
first place, $T_H$ would be independent of $B$ and would be the usual
closed string value $1/4\pi\sqrt{\alpha'}$.  This non-commutativity of
compactifying and going to the Hagedorn temperature is a result of the
exponential growth of the density of states of the string which is
independent of compactification radius.At the Hagedorn temperature, the 
thermal distribution of string states is unstable and the most favorable
configuration is one long string that contains all of the energy.  In order
to know about the $B$-field, this long string must wrap the compactified
light-like direction.  Because of this non-extensive behavior, it always
has enough energy to do that, no matter how large the radius $R$.

In Section 2 we shall examine the free energy of a Bosonic string in a
background $B$-field.  We will use discrete light-cone quantization (DLCQ)
and compute the thermodynamic partition function using operator quantization
in the light-cone gauge.  

In Section 3, we compute the same partition function using the
covariant path integral. We do this to illustrate a peculiarity of the
compactifications that must be implemented in our computation.  It is
necessary to do a simultaneous compactification of a null direction in
Minkowski space in order to get DLCQ and Euclidean time, in order to
get a finite temperature path integral.  We shall see that the complex
identification of time, contained in formulae (\ref{lcc}) and
(\ref{x0c}) indeed produces a partition function which agrees with the
one obtained by operator methods.

In Section 4, we use this technique to examine the effect of these
simultaneous compactifications on the string path integral at all
genera.  We show how they lead to a constraint in the path integral
measure which restricts the integration over all Riemann surfaces to
an integration over the moduli space of branched covers of a
particular torus.  We find that the external $B$-field enters the
partition function only in these constraints.

In Section 5 we extend our genus 1 results for the bosonic string to
the type IIA superstring.  We obtain a formula for the genus 1
contribution to the thermodynamic free energy of the DLCQ superstring
in a constant $B$-field quoted in (\ref{supfe1}).

In Section 6, we examine the thermodynamic partition function of
Matrix string theory in a $B$-field.  We first identify the coupling
of a constant $B$-field to the Matrix model.  Then we examine the
matrix string free energy in the limit where we keep only the diagonal
components of all of the matrix fields.  We give an explicit
derivation of the thermodynamic partition function at genus 1.  The
result is identical to the genus 1 superstring partition function that
we derived in Section 5.

In Section 7, we discuss the Hagedorn temperature in the discrete
light-cone quantized system.  We note a peculiarity of the Hagedorn
temperature.  It depends on the $B$-field, but not on the light-cone
compactification radius.  

In Section 8 we give some concluding remarks.

%\newpage
%\setcounter{page}1

\section{Free Energy of the Bosonic String in a B-field: DLCQ}

\subsection{Notation}
Let us begin by summarizing some of our notation and conventions.
We will use the metric of $D$-dimensional Minkowski space given by 
\begin{equation}
-ds^2=dx^\mu g_{\mu\nu} dx^\nu\equiv -(dx^0)^2+{d\vec x}^{~2}
=-2dx^+ dx^- +\left( d\vec x_T\right)^2
\end{equation}
where $\vec x$ is the vector made from the $D-1$ spatial components of
$x^\mu$, the light-cone coordinates are
\begin{equation}
x^+=\frac{1}{\sqrt{2}}\left(x^0+x^1 \right)
~~,~~~
x^-=\frac{1}{\sqrt{2}}\left(x^0-x^1 \right)
\end{equation}
and  
$
\vec x_T=\left(x^2,\ldots,x^{D-1}\right)
$. We will always consider the string in critical dimensions, so that
$D=26$ for the Bosonic string and $D=10$ for the superstring.

In Euclidean space
$$
X^0 \to -iX^0_E ~~~,~~ X^i \to X^i
$$
consequently
$$
B_{01} \to iB_{01}^E
$$

The closed string worldsheet coordinates are denoted by $\tau$ and 
$\sigma$ where 
$\sigma\in[0,\pi]$ as in \cite{gsw}. We define
$$
\sigma_{\pm}= \frac{1}{\sqrt{2}}\left(\tau \pm \sigma \right)
$$

$$
\partial_{\pm}= \frac{1}{\sqrt{2}}
\left(\partial_\tau \pm \partial_\sigma \right)
$$
When both the target space and world-sheet are Euclidean
$$
\tau \to -i\sigma_2 ~~~,~~ \sigma \to \sigma_1
$$

The four Jacobi theta functions that we will use are defined by
\begin{eqnarray}
\theta_1(\nu,\tau)=i\sum_{n=-\infty}^\infty (-1)^nq^{(n-1/2)^2/2}z^{n-1/2}
~,~
\theta_3(\nu,\tau)=\sum_{n=-\infty}^\infty q^{n^2/2}z^{n}
\\
\theta_2(\nu,\tau)=\sum_{n=-\infty}^\infty q^{(n-1/2)^2/2}z^{n-1/2}
~,~
\theta_4(\nu,\tau)=\sum_{n=-\infty}^\infty (-1)^nq^{n^2/2}z^{n}
\end{eqnarray}
where 
$$
q=\exp\left(2\pi i\tau\right)~~~,~~~z=\exp\left(2\pi i\nu\right)
$$
We shall also denote by $$\tilde
\theta_k(\nu,\tau)=\sum_{\stackrel{n=-\infty}
{n\neq0}}^\infty \ldots$$ 
the theta function where the $n=0$ term is absent from the sum.

Under the two generators of the $SL(2,Z)$ modular group, $\tau\to\tau+1$ and
$\tau\to -1/\tau$, 
the modular transformation properties of the theta functions are
\begin{eqnarray}
\theta_1(\nu,\tau+1)=e^{i\pi/4}\theta_1(\nu,\tau)~~&,&~~
\theta_2(\nu,\tau+1)=e^{i\pi/4}\theta_2(\nu,\tau)\cr
\theta_3(\nu,\tau+1)=\theta_4(\nu,\tau)~~&,&~~
\theta_4(\nu,\tau+1)=\theta_3(\nu,\tau)
\label{modular1}
\end{eqnarray}
and
\begin{eqnarray}
\theta_1(\nu/\tau, -1/\tau)=-\left(-i\tau\right)^{1/2}e^{\pi i\nu^2/\tau}
\theta_1(\nu,\tau) \cr
\theta_2(\nu/\tau, -1/\tau)=-\left(-i\tau\right)^{1/2}e^{\pi i\nu^2/\tau}
\theta_4(\nu,\tau) \cr
\theta_3(\nu/\tau, -1/\tau)=-\left(-i\tau\right)^{1/2}e^{\pi i\nu^2/\tau}
\theta_3(\nu,\tau) \cr
\theta_4(\nu/\tau, -1/\tau)=-\left(-i\tau\right)^{1/2}e^{\pi i\nu^2/\tau}
\theta_2(\nu,\tau) 
\label{modular2}
\end{eqnarray}
They obey Jacobi's abstruse identity
\begin{equation}
\theta_3^4(0,\tau)-\theta_4^4(0,\tau)-\theta_2^4(0,\tau)=0
\label{abstruse}
\end{equation}

Jacobi's triple product formulae are
\begin{eqnarray}
\theta_1(\nu,\tau)=2q^{1/8}\sin(\pi\nu)\prod_{n=1}^\infty
\left(1-q^n\right)\left(1-2q^n\cos(2\pi\nu)+q^{2n}\right) \cr
\theta_2(\nu,\tau)=2q^{1/8}\cos(\pi\nu)\prod_{n=1}^\infty
\left(1-q^n\right)\left(1+2q^n\cos(2\pi\nu)+q^{2n}\right) \cr
\theta_3(\nu,\tau)=\prod_{n=1}^\infty
\left(1-q^n\right)\left(1+2q^{n-1/2}\cos(2\pi\nu)+q^{2n-1}\right) \cr
\theta_4(\nu,\tau)=\prod_{n=1}^\infty
\left(1-q^n\right)\left(1-2q^{n-1/2}\cos(2\pi\nu)+q^{2n-1}\right) \cr
\label{triple}
\end{eqnarray}

The Dedekind eta function is
\begin{equation}
\eta(\tau)=q^{1/24}\prod_{n=1}^\infty\left( 1-q^n\right)
\label{eta}
\end{equation}
and has the modular transformation properties
\begin{equation}
\eta(\tau+1)=e^{i\pi/12}\eta(\tau)~~,~~\eta(-1/\tau)=(-i\tau)^{1/2}
\eta(\tau)
\end{equation}

We shall also use the notation
\begin{equation}
\bar{\eta}(\tau)=\bar{q}^{1/24}\prod_{n=1}^\infty\left( 1-\bar{q}^{n}\right)
\label{etabar}
\end{equation}
where $\bar{q}=e^{-2\pi i\bar{\tau}}$.

The Poisson re-summation formula is
\begin{equation}
\sum_{n=-\infty}^{\infty}\exp{\left(-\pi an^2 -2\pi ibn\right)}
=a^{-1/2}\sum_{m=-\infty}^{\infty}\exp{\left[
-\frac{\pi (m+b)^2}{a}\right]}
\label{resum}
\end{equation}

\subsection{Action, Equations of Motion and Solutions}

The action for the closed Bosonic string on 26-dimensional Minkowski space
in the presence of a $B$-field is 
given by
\begin{equation}
S=
-\frac{1}{4\pi\alpha'}\int d^2 {\sigma}\left(
\partial_{\alpha} X^\mu \partial^{\alpha} X_\mu -2\pi\alpha' B_{\mu\nu}
\epsilon^{\alpha\beta}\partial_{\alpha} X^\mu \partial_{\beta} X^\nu\right)
\label{c1}
\end{equation}
Here we have fixed the conformal gauge for the worldsheet metric.
When $B_{\mu\nu}$ is a constant, the term in the world-sheet Lagrangian 
density containing it is a total derivative. For this reason, this term 
does not alter the equations of motion, which will therefore be 
independent of $B$. 
These equations of motion and constraints are 
\begin{eqnarray}
\partial_+\partial_- X^\mu =0 \\
\partial_+X^\mu\partial_+ X_\mu=0 \\
\partial_-X^\mu\partial_- X_\mu=0
\label{c3}
\end{eqnarray}
The conserved world-sheet 
Noether currents which are associated with space-time translation
invariance depend on $B$, 

\begin{equation}
{\cal P}^\mu_{\alpha} = \frac{1}{2\pi \alpha '}\partial 
_{\alpha}X^\mu -{B^\mu_{~\nu}}\epsilon_{\alpha}^{\ \beta}\partial_{\beta}X^\nu
\label{c18}
\end{equation}

In DLCQ a null direction is compactified.  We shall consider the
case where 
$X^+$ is identified with $X^++2\pi R$. 
In closed string theory, the boundary conditions on the worldsheet
are periodic,
\begin{equation}
X^\mu(\tau,\sigma+\pi)=X^\mu(\tau,\sigma)+\delta^\mu_+2\pi Rr
\label{bc}
\end{equation}
where $r$ is an integer which counts the number of times 
the string world-sheet wraps the compact direction $X^+$. 
When the string wraps the compact direction,
the Noether charges are influenced by $B_{\mu\nu}$ as
\begin{equation}
P^\mu=\frac{1}{2\pi\alpha'}\int_0^\pi d\sigma \partial_\tau X^\mu 
+ 2\pi r B^\mu_{~+} 
\end{equation}
We will consider the situation where the only non-zero component of $B$ is $B_{+-}$.
In that case, the only momentum which is influenced by $B$ is $P^+$ and 
\begin{equation}
P^+=\frac{1}{2\pi\alpha'}\int_0^\pi d\sigma \partial_\tau X^+ + 2\pi B r
\end{equation}
The equations of motion have the light-cone gauge solutions
\begin{eqnarray}
X^-(\tau,\sigma)&=&x^-+2{\alpha'}P^-\tau  \label{x-}\quad\quad\\
X^+(\tau,\sigma)&=&x^+ +2 \alpha' (P^+-2\pi B r) \tau +2Rr\sigma 
+\\
&+&\left(\frac{\alpha'}{2}\right)^{(1/2)}i\sum_{n\neq 0}
\left(\frac{\alpha _n^+}{n}e^{-2in\left(\sigma +\tau\right)}+
\frac{\tilde{\alpha}_n^+}{n}e^{+2in\left(\sigma -\tau\right)} \right)
\label{x+}\quad\quad\\
\vec X_T(\tau,\sigma)&=&\vec x_T +{2 \alpha'}\vec{P}_T\tau 
+\left(\frac{\alpha'}{2}\right)^{(1/2)}i\sum_{n\neq 0}
\left(\frac{\vec{\alpha}_n}{n}e^{-2in\left(\sigma +\tau\right)}+
\frac{\vec{\tilde{\alpha}}_n}{n}e^{+2in\left(\sigma -\tau\right)} \right)
\quad\quad
\label{xt}
\end{eqnarray}
where $\vec{P}_T$ and $P^-$ are the total momentum along the
transverse and $X^-$ directions, respectively.  As we shall show, in
the presence of a $B$-field, $p^+$ does not
coincide with the translation generator along the $X^+$ direction,
which we will denote by $P^+$.

Since the canonical commutation relations are
$$
\left[x^+,P^-\right]=-i
$$
and $x^+$ is a compact variable, the momentum $P^-$ is quantized as
$$
P^-= \frac{k}{R}
$$
where $k$ is any integer.
Substituting the solution (\ref{x-}, \ref{x+}, \ref{xt}) into 
the equations of motion, when $k\neq 0$, we get 
the explicit solutions for the $\alpha_n^+$ in terms of
the transverse oscillators $\alpha^i_n$
\begin{eqnarray}
{\alpha}_n^+ &=& \frac{1}{2P_-}\left(\frac{2}{\alpha '}\right)^{1/2}
\sum_{m=-\infty}^{+\infty}:\alpha _{n-m}^{i}\alpha _{m}^{i}: -
\frac{1}{P_-}\left(\frac{2}{\alpha '}\right)^{1/2}\delta_{n,0}  \cr
{\tilde{\alpha}}_n^+ &=& \frac{1}{2P_-}\left(\frac{2}{\alpha '}
\right)^{1/2}\sum_{m=-\infty}^{+\infty}:\tilde{\alpha}_{n-m}^{i}
\tilde{\alpha}_{m}^{i}: -
\frac{1}{P_-}\left(\frac{2}{\alpha '}\right)^{1/2}\delta_{n,0}
\label{alpha0}
\end{eqnarray}
where $\alpha^\mu_0=p^\mu$.
Subtracting these equations for $n=0$ we get
\begin{equation}
P^- r R= k r=N-\tilde N
\end{equation}
where 
\begin{equation}
N=\sum_{n=1}^\infty\alpha_{-n}^i\alpha_n^i~,~~~
\tilde{N}=\sum_{n=1}^\infty\tilde{\alpha}_{-n}^i\tilde{\alpha}_n^i\ ,
\label{levels}
\end{equation}
This is the level-matching condition, $L_0 -\tilde{L}_0 =0$.

Using the solutions (\ref{alpha0}) we get
\begin{equation}
P^+= \frac{1}{\alpha 'P^-}\left(N+\tilde{N}-2\right)
+\frac{1}{2P^-}P^i P^i 
+\frac{2\pi B}{P^-}\left(N-\tilde{N}\right)
\end{equation}
When we take into account the level matching condition, $N-\tilde
N=kr$, we see that this agrees with the expression (\ref{boostedmom})
which we found by considering a compactified spatial direction and
doing an infinite boost in that direction.

The above  relation gives the mass-shell condition
\begin{eqnarray}
M^2 &=& 2P^+ P^- -P^i P^i \cr
&=&\frac{2}{\alpha'}\left( N+\tilde N-2\right)+4\pi B\left(N-\tilde{N}\right)
\label{masscond}
\end{eqnarray}
This coincides with (\ref{tensionshift}).

\subsection{Thermodynamic Free Energy}

From ref.\cite{Grignani:1999sp} it is now easy to construct the free
energy of the bosonic string in the presence of a constant $B$-field.
The only difference lies in the mass spectrum operator which is now
given by (\ref{masscond}).  We use an expression for the free energy
of a single relativistic particle and sum it over the mass spectrum of
the string.  The result is
\begin{eqnarray}
\frac{F}{V}=-\sum_{n=1}^\infty \sum_{k=1}^\infty \frac{1}{k} 
\left( \frac{k}{\sqrt{2}\pi n\beta R}\right)^{13}~
\sum_{s=0}^{k-1}\frac{1}{k}
{\rm  Tr}~
\exp\left(-\frac{n\beta R}{\sqrt{2}}
\frac{M^2}{2k}-\frac{nk\beta}{\sqrt{2}R}
\right. \cr \left.
+2\pi i 
\frac{s}{k}(\tilde{N}-N)\right)
\label{42}
\end{eqnarray}
where ${\rm Tr}$ denotes the sum over string states with $M^2$,
$N$ and $\tilde N$ the operators given in (\ref{masscond}) and
(\ref{levels}).  In (\ref{42}), summation over the integer $s$
enforces the level matching condition.  The summation over $k$ comes
from expanding the free energy of the relativistic particle in a
series of exponentials. The summation over $n$ is the sum over
light-cone momenta dual to the compact null direction.  Some details
about how this formula is obtained can be found in
\cite{Grignani:1999sp}.  We use the notation for $\tau_\pm$, $\hat\tau_2$ and
$\nu$ given in (\ref{tauplus}), (\ref{tauminus}), (\ref{tautwo}) and (\ref{nu}).
In terms of these parameters, the free energy of the Bosonic string
is given by
\begin{eqnarray}
\frac{F}{V}=-\sum_{n=1}^\infty \sum_{k=1}^\infty \sum_{s=0}^{k-1}
\frac{1}{k^2} \left(\frac{1}{4\pi^2\alpha'\hat{\tau}_2}\right)^{13}
e^{4\pi\hat{\tau}_2-n^2\beta^2/4\pi\alpha'\hat{\tau}_2}
{\rm Tr}\left(e^{-2\pi i\bar{\tau}_+ N}~e^{2\pi i \tau_- \tilde N}\right)
\label{fb}
\end{eqnarray}
The trace over the string states can be easily computed 
and reads
\begin{equation}
\frac{F}{V}=-\sum_{n=1}^\infty \sum_{k=1}^\infty \sum_{s=0}^{k-1}
\frac{1}{k^2} \left(\frac{1}{4\pi^2\alpha'\hat{\tau}_2}\right)^{13}~
\exp\left(-\frac{n^2\beta^2}{4\pi\alpha'\hat{\tau}_2}\right)
\left(\eta\left(\tau_-\right)\right)^{-24}
\left(\bar{\eta}\left(\tau_+\right)\right)^{-24}
\label{fb1}
\end{equation}
where the Dedekind eta function is defined in equation
(\ref{eta}).  This is the free energy density of the DLCQ bosonic
string in a constant $B$-field. For $B=0$, equation (\ref{fb1})
exactly reproduced the result obtained in ref. \cite{Grignani:1999sp}.

We shall show in the next section how
this free energy can be obtained using the covariant 
Polyakov path integral.

\section{Covariant Path Integral}
In this section we will calculate the free energy of the bosonic string 
in the presence of a $B$-field at genus 1 using the covariant approach.
We will assume that the metrics of both the world-sheet and the target 
spacetime have Euclidean signatures.
The action in Euclidean space is given by
\begin{equation}
S=\frac{1}{4\pi\alpha'}\int d^2\sigma \left( \sqrt{g}g^{\alpha \beta}
\partial_{\alpha}X^\mu\partial_{\beta}X^\mu -2\pi\alpha'iB^E_{\mu\nu}
\epsilon^{\alpha \beta}
\partial_{\alpha}X^\mu \partial_{\beta}X^\nu \right)
\label{act}
\end{equation}
Note that we have also Wick rotated the antisymmetric tensor fields, 
$B_{01}\to + iB^E_{01}$.  We shall show that the compactifications of 
the target space lead to a well-defined path integral 
also when $B^E$ is complex.  To achieve this result we shall use
Seiberg approach to the DLCQ of the bosonic string~\cite{Seiberg:1997ad}:
We shall obtain a compactified null direction from a compactified
spatial direction (given by a small circle of radius $R_s$)
by an infinite boost along the compact direction.
At the end of the computation we shall remove the cutoff $R_s$
and we shall analytically continue back to Minkowski $B$.
The cutoff $R_s$ is introduced to show which is the correct
prescription to give a path integral representation of the DLCQ
free energy. Namely we shall show that the result obtained using 
Seiberg's approach to DLCQ can be gotten directly
compactifying a null direction ($i.e.~R_s=0$) but with a 
$real$ Euclidean $B^E$,  
which only in the final equation is replaced by a Minkowskian $B$.
This is the procedure we shall adopt in the rest of the Paper.

The spacetime is assumed to be flat 26-dimensional Euclidean space.
The genus 0 Riemann surface is topologically the 2-sphere.  Since it has
no non-contractible loops, it cannot depend on $B^E$, $\beta$ or $R$.  

At genus 1 the world-sheet is a torus.
We will take the coordinates of the world-sheet, $\sigma^1$ and $\sigma^2$ 
to lie in the range $\sigma^i\in[0,1)$.  Conformal transformations can always
be used to put the metric of the torus in the form
\begin{equation}
g_{\alpha \beta}=\left( \matrix{ 1 & \tau_1 \cr \tau_1 & 
|\tau |^2 \cr }\right)
\label{wsmetric}
\end{equation}
where then entries are constants given by the complex Teichm\"uller
parameter
\begin{equation}
\tau=\tau_1+i\tau_2
\end{equation}
The determinant of the metric is
\begin{equation}
|g|=\tau_2^2
\end{equation}
and the inverse of the metric is
\begin{equation}
g^{\alpha \beta}=\frac{1}{\tau_2^2}\left( \matrix{ |\tau|^2 & -\tau_1\cr
 -\tau_1 & 1 \cr }\right)
\end{equation} 
With this gauge fixing of the metric,
the integration over the metric in the Polyakov path integral becomes
an integration over the Teichm\"uller parameter.

We shall also assume that $B^E_{\mu \nu}$ is a constant. The factor
$i$ in front of the term containing $B^E_{\mu\nu}$ comes from analytic
continuation.  The contribution of the $B$-field term to the
world-sheet Lagrangian density is proportional to a total derivative
and therefore it depends only on the topology of the configuration.
We wish to study the situation where the target space has particular
compact dimensions.  In this case, string configurations which wrap
the compact dimensions have non-trivial topology.

Two compactifications will be needed.  The first compactifies the
light cone Minkowski space by making the identification
$\frac{1}{\sqrt{2}}(t+x^{1})\sim
\frac{1}{\sqrt{2}}(t+x^{1})
+2\pi R$. In our Euclidean coordinates it becomes the complex
identification
\begin{equation}
\left(X^0,X^1,\vec{X}_T\right)\sim
\left(X^0+\sqrt{2}\pi i R,X^1+\sqrt{2}\pi R,\vec{X}_T\right)
\label{lcc}
\end{equation}
The second compactification that we shall need is that of
Euclidean time which is necessary to
produce finite temperature $T=1/\beta$, 
\begin{equation}
\left(X^0,X^1,\vec{X}_T\right)\sim
\left(X^0+\beta,X^1,\vec{X}_T\right)
\label{x0c}
\end{equation}
With these compactifications the only relevant component of the
$B^E_{\mu \nu}$ tensor is $B^E_{01}=-B^E_{10}=B^E$.

Following Seiberg~\cite{Seiberg:1997ad} we shall consider the
light-like compactification as the limit of a compactification on a 
space-like circle which is almost light-like
\begin{equation}
\left(X^0,X^1,\vec{X}_T\right)\sim
\left(X^0+\sqrt{2}\pi i R,X^1+\sqrt{2}\pi R\sqrt{1+
2\left(\frac{R_s}{R}\right)^2},\vec{X}_T\right)
\label{lcc1}
\end{equation}
with $R_s<<R$. The light-like circle (\ref{lcc}) is obtained from 
(\ref{lcc1}) as $R_s\to 0$. This compactification  is related by
a large boost with
$$v=\frac{R}{\sqrt{R^2+2R^2_s}}$$
to a spatial compactification on
$$\left(X^0,X^1,\vec{X}_T\right)\sim
\left(X^0,X^1+{2}\pi R_s,\vec{X}_T\right)$$
The introduction of the cutoff $R_s$, which will be removed 
at the end of the calculation, is used to make the
analytic continuation of the field $B^E$ to the Minkowski
$B$, well defined at any stage of the computation.

Compactification is implemented by including the possible wrappings of 
the string world-sheet on the compact dimensions.
In general
the bosonic coordinates of the string should have a multi-valued part 
which should take into account the following boundary conditions
\begin{eqnarray}
X_T^i\left(\sigma_1 +1 , \sigma_2 \right)&=&X_T^i\left(\sigma_1 , 
\sigma_2 \right) \cr
X_T^i\left(\sigma_1 , \sigma_2 +1 \right)&=&X_T^i\left(\sigma_1 , 
\sigma_2 \right) \cr
X^0\left(\sigma_1 +1 , \sigma_2 \right)&=&X^0\left(\sigma_1 , 
\sigma_2 \right) +\beta m+i\sqrt 2 \pi Rp \cr
X^0\left(\sigma_1 , \sigma_2 +1 \right)&=&X^0\left(\sigma_1 , 
\sigma_2 \right) +\beta n+i\sqrt 2 \pi Rq \cr
X^{1}\left(\sigma_1 +1 , \sigma_2 \right)&=&X^{1}\left(\sigma_1 , 
\sigma_2 \right) +\sqrt 2 \pi R' p \cr
X^{1}\left(\sigma_1 , \sigma_2 +1 \right)&=&X^{1}\left(\sigma_1 , 
\sigma_2 \right) +\sqrt 2 \pi R' q
\label{boundcs}
\end{eqnarray}
where $R'=R\sqrt{1+2\left(\frac{R_s}{R}\right)^2}$.

As it was shown in Refs.~\cite{O'Brien:1987pn}~\cite{McClain:1987id}
with the boundary conditions (\ref{boundcs})
conformal transformations can be used to transform the integration 
region over the Teichm\"uller parameter $\tau$ into
the fundamental domain
\begin{equation}
{\cal F}=\left\{ z\in{\cal C} | |z|\geq 1 , -1/2\leq {\rm re}z < 1/2
\right\}
\label{fundamental}
\end{equation}
To arrive at the form of the bosonic free energy we obtained
in the previous section, eq.(\ref{fb1}), however, we should use 
modular transformations to characterize the string 
wrappings by a single integer~\cite{O'Brien:1987pn}
\cite{McClain:1987id}.  In doing so, the integration domain for 
the Teichm\"uller parameters 
is expanded from ${\cal F}$ to the region 
\be
{\cal S}\equiv\left\{-1/2<\tau_1<1/2,\   0<\tau_2<\infty\right\}
\label{cals}
\ee
As was shown in the seminal Paper by Polchinski \cite{Polchinski:1986zf}, 
this corresponds, on the torus, to consider
only the windings of the $time$ coordinate of the torus, $\sigma_2$,
around the target space circle in the $X^0$ direction. 
In the Seiberg approach to DLCQ this means boundary conditions of the form
\begin{eqnarray} 
X^0\left(\sigma_1 +1 , \sigma_2 \right)&=&X^0\left(\sigma_1 ,  
\sigma_2 \right) +i\sqrt 2 \pi R p \cr 
X^0\left(\sigma_1 , \sigma_2 +1 \right)&=&X^0\left(\sigma_1 ,  
\sigma_2 \right) +\beta n+i\sqrt 2 \pi Rq \cr 
X^{1}\left(\sigma_1 +1 , \sigma_2 \right)&=&X^{1}\left(\sigma_1 ,  
\sigma_2 \right) +\sqrt 2 \pi R'p \cr 
X^{1}\left(\sigma_1 , \sigma_2 +1 \right)&=&X^{1}\left(\sigma_1 ,  
\sigma_2 \right) +\sqrt 2 \pi R'q 
\label{boundcs1} 
\end{eqnarray} 
which coincide with (\ref{boundcs}) for $m=0$.
We can then write the embedding coordinates $X^{0},\ X^{1}$, 
as a periodic part plus a multi-valued part, according to
\begin{eqnarray}
X^0&=&X^0_p +n\sigma_2 \beta +\sqrt 2 i\pi R
\left(p\sigma_1 +q\sigma_2 \right) \cr
X^{1}&=&X^{1}_p +\sqrt 2 \pi R'\left(p\sigma_1 +q\sigma_2 \right)
\label{coord}
\end{eqnarray}
With these identifications of the spacetime the action (\ref{act}) 
can be written as
\begin{equation}
S=\frac{1}{4\pi\alpha'}\int d^2\sigma \sqrt{g}g^{\alpha \beta}
\partial_{\alpha}X^\mu _p \partial_{\beta}X^\mu _p + S_{w.m.}(n,p,q)
\label{action}
\end{equation}
where we separated the contributions coming from the periodic part
from those due to the winding modes. Note that $\vec{X_p}_T\equiv
\vec{X}_T$. Plugging (\ref{coord}) into (\ref{action}) for 
$S_{w.m.}$ one gets
\be
S_{w.m.}(n,p,q)=\frac{\beta^2 n^2}{4\pi \alpha '\tau_2}+
\frac{\pi R_s^2}{\alpha '\tau_2}\left|p\tau -q\right|^2
+ \frac{2\pi i}{\nu\tau_2}
\left(nq
-(\tau_1 -2\pi\alpha'B^E \frac{R'}{R}\tau_2)np\right)
\label{swm}
\ee
where $\nu=4\pi\alpha'/(\sqrt{2}\beta R)$.
The part of the path integral which depends on the topological data is
gotten by exponentiating the action (\ref{swm}) and summing over the
integers $n,p,q$. 
The expression for the free energy becomes
\begin{equation}
\frac{F}{V}=-\sum_{n,p,q}\int_{\cal S} d\tau_1 \frac{d\tau_2}{2\tau_2} 
\frac{\left| \eta
\left(\tau\right)\right|^{-48}}{(4\pi^2\alpha'\tau_2)^{13}}
e^{-S_{w.m}(n,p,q)}
\label{fe}
\end{equation}
where the limit $R_s\to 0$ is omitted and the Dedekind eta function
and the powers of $4\pi^2\alpha'\tau_2$ arise from the integration
over $X^\mu_p$ and gauge fixing.
The limit $R_s\to 0$ is not defined inside the integral for a 
complex $B^E$, in fact if $B^E$ has an imaginary part 
and we set $R_s=0$ inside the integral
the sums over $p$ and $q$ diverge, since $p$ and $q$ 
would appear linearly in 
(\ref{swm}). 
To take this limit, which
would provide the DLCQ free energy for the bosonic string,
it is convenient to first Poisson re-sum over $p$ and $q$,
then perform the integrals over $\tau_1$ and $\tau_2$ and only at the
end remove the cutoff by taking $R_s\to 0$. One gets
\be
\sum_{n,p,q}e^{-S_{w.m.}(n,p,q)}=\frac{\alpha'}{R^2_s}
\sum_{n,k,s}e^{-S'_{w.m.}(n,k,s)}
\ee
where
\be
S'_{w.m.}(n,k,s)=
\frac{\beta^2 n^2}{4\pi \alpha '\tau_2}+
\frac{\pi \alpha'}{R_s^2 \tau_2}\left[\left((s-\tau_1 k+
\frac{\sqrt{2}}{2} B^E\beta R'n\right)^2+\left(\tau_2 k -
\frac{n}{\nu}\right)^2\right]
\label{spwm}
\ee

The free energy becomes
\bea
\frac{F}{V}&=&-\sum_{n,k,s}\int^{\frac{1}{2}}_{-\frac{1}{2}} d\tau_1 
\int_0^\infty\frac{d\tau_2}{2\tau_2} 
\frac{\left|\eta(\tau)\right|^{-48}}{(4\pi^2\alpha'\tau_2)^{13}}
\frac{\alpha'}{R_s^2}
\exp\left[-\frac{\beta^2n^2}{4\pi \alpha '\tau_2}\right.\cr
& &\left. -\frac{\pi \alpha'}{R_s^2 \tau_2}\left(s-\tau_1 k+
\frac{\sqrt{2}}{2} B^E\beta R'n\right)^2-
\frac{\pi \alpha'}{R_s^2 \tau_2}\left(\tau_2 k - 
\frac{n}{\nu}\right)^2\right]
\label{fem0}
\eea
It is not difficult to show that, in the $R_s\to 0$ limit,
the $k=0$ term in this integral is different from zero only for $n=0$.
For $n=0$ one gets a temperature independent contribution
which gives the cosmological constant at one-loop 
\cite{Polchinski:1986zf}. Thus in what follows we can set $n\ne0$
and $k\ne 0$. In particular, since (\ref{fem0}) is symmetric for $k\to-k$,
$n\to-n$ and $s\to -s$ we can also choose $k=1,\dots\infty$
and multiply by a factor of 2.

To perform the integration over $\tau_1$ it is useful to 
rewrite the Dedekind eta function in terms of a series
as in \cite{gsw}. One has
\begin{equation}
\left|\eta(\tau)\right|^{-48}=e^{4\pi\tau_2}\left|\prod_{m=1}^\infty
\left(1-e^{2\pi i \tau m}\right)\right|^{-48}
\end{equation}
and  
\begin{equation}
\prod_{m=1}^\infty
\left(1-z^m\right)^{-24}\equiv \sum_{r=0}^\infty p(r) z^r
\end{equation}
where $z=\exp(2\pi i \tau)$.
Then the  $\tau_1$ dependent terms can be rewritten as 
\bea
\int^{\frac{1}{2}}_{-\frac{1}{2}} d\tau_1 
\exp\left[-\frac{\pi \alpha'}{R_s^2 \tau_2}\left(s-\tau_1 k+
\frac{\sqrt{2}}{2} B^E\beta R'n+i\frac{R_s^2 \tau_2}{k\alpha'}(r-r')
\right)^2 \right.\cr\left.
-\frac{R_s^2 \tau_2}{k^2\alpha'}(r-r')^2+2\pi i
\left(\frac{s}{k}+\frac{\sqrt{2}}{2k} B^E\beta R'n\right)(r-r')\right]
\label{tau1int}
\eea
To perform the $\tau_1$ integral it is useful to define in (\ref{fem0})
new summation variables. Instead of summing over $s$
one can sum over two integers $s'$ and $l$ which are defined according 
to
$$
s'= s+l k
$$
where $-\infty<l<\infty$ and $s'=0,\dots,k-1$. Introducing the 
new integration variable
$$
\tau_1'=\tau_1+l
$$
eq.(\ref{tau1int}) with the sum over $s'$ and $l$ becomes
\bea
\sum_{l=-\infty}^{\infty}\sum_{s'=0}^{k-1}
\int^{\frac{1}{2}+l}_{-\frac{1}{2}+l} d\tau'_1 
\exp\left[-\frac{\pi \alpha'}{R_s^2 \tau_2}\left(s'-\tau'_1 k+
\frac{\sqrt{2}}{2} B^E\beta R'n+i\frac{R_s^2 \tau_2}{k\alpha'}(r-r')
\right)^2 \right.\cr\left.
-\frac{R_s^2 \tau_2}{k^2\alpha'}(r-r')^2+2\pi i
\left(\frac{s'}{k}+\frac{\sqrt{2}}{2k} B^E\beta R'n\right)(r-r')\right]
\eea
It is now clear that the sum over $l$ can be used to make the $\tau_1'$
integral Gaussian so that the free energy becomes (with $s'\to s$)
\bea
&&\frac{F}{V}=- \sum_{n\ne 0}\sum_{k=1}^{\infty}
\sum_{s=0}^{k-1}\sum_{r,r'=0}^{\infty}\sqrt{\frac{\alpha'}{k^2 R^2_s}}
\int_0^\infty\frac{d\tau_2}{\sqrt{\tau_2} (4\pi^2\alpha'\tau_2)^{13}}
 p(r) p(r')\cr
&&\exp\left[-\frac{\beta^2n^2}{4\pi \alpha '\tau_2}+
4\pi\tau_2-2\pi\tau_2(r+r')-
\frac{R_s^2 \tau_2}{k^2\alpha'}(r-r')^2\right.\cr
&&\left.+2\pi i
\left(\frac{s}{k}+\frac{\sqrt{2}}{2k} B^E\beta R'n\right)(r-r')
-\frac{\pi \alpha'}{R_s^2 \tau_2}\left(\tau_2 k - 
\frac{\sqrt{2}\beta R}{4\pi\alpha'}n\right)^2\right]
\eea
The $\tau_2$ integral can be performed exactly in terms of a Bessel function
$K_{25/2}$, but it is much easier to perform at this stage the limit
$R_s\to0$ which is now well defined for any complex $B^E$. 
In fact, using the formula
$$
\lim_{\epsilon\to 0}\frac{1}{\epsilon\sqrt{\pi}}e^{-x^2/\epsilon^2}=\delta(x)
$$
with $\epsilon=R_s/\sqrt{\alpha'}$ the free energy reads
\bea
\frac{F}{V}=- \sum_{n,k=1}^{\infty}
\sum_{s=0}^{k-1}\frac{1}{k^2}
\int_0^\infty\frac{d\tau_2}{(4\pi^2\alpha'\tau_2)^{13}}
\delta\left(\tau_2-\frac{n}{k\nu}\right)
\exp\left[-\frac{\beta^2n^2}{4\pi \alpha '\tau_2}\right]\cr
\left[\eta\left(\frac{s}{k}+i\frac{n}{k\nu}(1-i2\pi\alpha'B^E)\right)
\bar{\eta}\left(\frac{s}{k}+i\frac{n}{k\nu}(1-i2\pi\alpha'B^E)\right)
\right]^{-24}
\label{fepi}
\eea
where $\nu=4\pi\alpha'/(\sqrt{2}\beta R)$.
This formula precisely coincides with the one obtained directly in the
operatorial light-cone formalism, eq.(\ref{fb1}), once the Euclidean 
$B^E$-field is analytically continued to Minkowski space $B^E\to -iB$. 
Note that this analytical continuation can  be 
performed at any stage of the calculation of this section, thus showing that 
what we have obtained is a consistent path integral representation 
of the free-energy of the bosonic string in DLCQ for any complex $B$.
As we shall show in Section 5 this path integral representation can 
also be used for the superstring in the NSR formalism.

The result (\ref{fepi}) can be obtained more straightforwardly
using directly the compactification of the light-like direction $x^+$,
namely with boundary conditions on the embedding coordinates
given by (\ref{boundcs1}) with $R_s=0$. When the action
$S_{w.m.}$ is exponentiated, since $p$ and $q$ would appear
linearly in this case, the summation over $p$ and $q$ leads to a
periodic delta function in the path integral measure.  This delta
function imposes a linear constraint on the Teichm\"uller parameter.
It is given by
\begin{equation}
\sum_{ks}\delta \left(\frac{n\tau_1 }{\nu\tau_2}
+\frac{2\pi \alpha 'B^E n}{\nu}+s\right) 
\delta \left(\frac{n}{\nu\tau_2}-k\right)   =
\sum_{ks}
\frac{\tau_2}{k^2}
\delta^2 \left(\tau -\frac{ s+\tau_0n}{k}\right)
\label{delta}
\end{equation}
where the sum over integers $k$ and $s$ makes the Dirac delta functions
periodic and 
$$
\tau_0
=i \frac{ 1-2\pi\alpha' iB^E }{\nu}
$$
is the parameter defined in (\ref{tauzero}).

Since we have assumed that $\tau_1$ and $\tau_2$ in 
eq.(\ref{delta}), 
$B^E$ cannot be  analytically continued 
to Minkowski space at this stage.
This operation
can only be performed at the end of the calculation.
The expression for the free energy becomes
\begin{equation}
\frac{F}{V}=-\sum_{n,s,k}\int_{\cal S} d\tau_1 \frac{d\tau_2}{2k^2} 
\frac{\left| \eta
\left(\tau\right)\right|^{-48}}{(4\pi^2\alpha'\tau_2)^{13}}
e^{-\frac{\beta^2 n^2}{4\pi \alpha '\tau_2}}
\delta^2 \left(\tau -\frac{s +\tau_0 n}{k}\right)
\label{fedlcq}
\end{equation}
It is now easy to perform the integrals in 
$\tau_1$ and $\tau_2$. After the integrations one can sensibly
continue the result to Minkowski space $B^E \to-iB$.  
The free energy one obtains from (\ref{fedlcq}) is identical 
to (\ref{fb1}).

The prescription is now clear. One can do any calculation
directly using the compactified light-like direction, provided
the Euclidean $B^E$ field is real. The correct result for 
the free energy is obtained when the analytic continuation to
a Minkowski $B$-field is performed only at the end of the calculation.
That this procedure is correct has been proven by using Seiberg's 
approach to DLCQ where the analytic continuation to Minkowski
$B$ can be done at any stage of the calculation.

As a further application of this procedure we shall now
derive the free energy with a Teichm\"uller
parameter defined in the fundamental domain ${\cal F}$. 
The boundary conditions we shall assume 
on the embedding coordinates
are given in (\ref{boundcs}) with $R_s=0$ so that we can write
$X^{0},\ X^{1}$, 
as a periodic part plus a multi-valued part, according to
\begin{eqnarray}
X^0&=&X^0_p +\left(m\sigma_1 +n\sigma_2 \right)\beta +\sqrt 2 i\pi R
\left(p\sigma_1 +q\sigma_2 \right) \cr
X^{1}&=&X^{1}_p +\sqrt 2 \pi R\left(p\sigma_1 +q\sigma_2 \right)
\label{coord1}
\end{eqnarray}
With (\ref{coord1}), 
$S_{w.m.}$ in (\ref{action}) becomes
\begin{eqnarray}
S_{w.m.}(m,n,p,q)=\frac{\beta^2}{4\pi \alpha '\tau_2}\left(m\bar\tau -n\right)
\left(\tau m-n\right)\cr 
+ 2\pi i\frac{\sqrt 2 \beta R}{4\pi \alpha '\tau_2}
\left[nq+|\tau |^2mp 
-\tau_1(mq+np)\right]-2\pi i\sqrt 2 \beta R\frac{B^E}{2}(mq-np)
\label{swmm}
\end{eqnarray}
The free energy reads
\begin{equation}
\frac{F}{V}=-\sum_{n,m,p,q}\int_{\cal F} d\tau_1 \frac{d\tau_2}{2\tau_2} 
\frac{\left| \eta
\left(\tau\right)\right|^{-48}}{(4\pi^2\alpha'\tau_2)^{13}}
e^{-S_{w.m}(m,n,p,q)}
\label{fem}
\end{equation}
where the Teichm\"uller parameter is integrated over the region ${\cal F}$, 
defined in (\ref{fundamental}).
In this expression the sum over $m$ and $n$ is the sum over maps of the torus 
in which the $space$ and $time$ time coordinates, respectively, of the torus  
wrap the target space $S^1$ $m$ and $n$ times. 
Since $p$ and $p$ appear
linearly in (\ref{swmm}), the summation over $p$ and $q$ in (\ref{fem})
leads to a
periodic delta function in the path integral measure.  This delta
function imposes a linear constraint on the Teichm\"uller parameter.
It is given by
\begin{eqnarray}
\sum_{ks}\delta \left(\frac{|\tau |^2 m-n\tau_1 }{\nu\tau_2}
+\frac{2\pi \alpha 'B^E n}{\nu}+s\right) 
\delta \left(\frac{n-m\tau_1}{\nu\tau_2}
-\frac{2\pi \alpha 'B^E m}{\nu} -k\right) \cr   =
\sum_{ks}\frac{\tau_2}{[(m 2\pi\alpha' B^E/\nu+k)^2+(m/\nu)^2]}
\delta ^2 \left(\tau -\frac{s +\tau_0 n}{k +\tau_0 m}\right)
\label{deltam}
\end{eqnarray}
where the sum over integers $k$ and $s$ makes the Dirac delta functions
periodic. This formula will be generalized to higher genera in the next 
section.
The discrete Teichm\"uller parameter then reads
\begin{eqnarray}
\tau(n,k,m,s)&=&\frac{s +\tau_0  n}
{k +\tau_0 m} \cr
\tau_1(n,k,m,s)&=& \frac{nm+\nu^2(n\phi/2\pi+s)
(m\phi/2\pi-k)}{m^2+\nu^2(m\phi/2\pi+k)^2}\cr
\tau_2(n,k,m,s)&=&\frac{\nu (nk-ms)}{m^2+\nu^2(m\phi/2\pi+k)^2}
\label{dtpm}
\end{eqnarray}
where
$$
\phi=\sqrt{2}\pi\beta R B^E
$$
is the flux of the $B^E$-field on the toroidal space.
Integrating over $\tau_1$ and $\tau_2$ the expression for the free 
energy becomes
\begin{eqnarray}
\frac{F}{V}=-\sum_{n,m,k,s}
\frac{1}{[(m\phi/2\pi+k)^2+(m/\nu)^2]} \left(\frac{1}{4\pi^2\alpha'\tau_2}
\right)^{13}~
\exp\left(-\frac{|m\tau-n|^2\beta^2}{4\pi\alpha'\tau_2}\right)
\left| \eta\left(\tau\right)\right|^{-48}\cr
\label{fb3}
\end{eqnarray}
Here, the integers $n,m,k,s$ must be chosen so that $\tau(n,k,m,s)$
lies in the fundamental domain ${\cal F}$ defined in
(\ref{fundamental}).

In equation (\ref{fb3}), the periodicity of the free energy when
$\phi\to\phi+2\pi \cdot{\rm integer}$, is manifest. Moreover when
$\phi=2\pi \cdot{\rm integer}$ the $B$-field contribution to the
partition function can be eliminated.  This is consistent with the
fact that this particular $B^E$-field can be eliminated by a gauge
transformation.

When $B^E\to - iB$ one obtains from (\ref{fb3}) the free energy
in a physical $B$-field which is related to (\ref{fb1})
by a modular transformation.

\section{Higher Genera}

The arguments of Section 3 can readily be extended to higher 
order perturbative contributions to the free energy of the string.  
These are obtained from the
covariant path integral where the worldsheets have higher genus.
We shall adopt the procedure described in the previous section
namely we shall perform the calculations in DLCQ
with a compactified light-like direction and a real Euclidean $B^E$-
field and only at the end of the calculation we shall analytically 
continue the result to Minkowski space.

\subsection{Notation}

We shall assume that the world-sheet is 
a Riemann surface $\Sigma_g$ of genus $g$.
We  use complex coordinates  
$$
z=\sigma_1+i\sigma_2~,~~~~~~~~~~~\bar z=\sigma_1-i\sigma_2
$$
Accordingly
$$
\partial_1=\partial_z+\partial_{\bar z}~,~~~~~~~~~~~~
\partial_2=i\left(\partial_z-\partial_{\bar z}\right)
$$
A local basis for the differential forms of the second order is given by
$d^2\sigma=
d\sigma_1\wedge d\sigma_2$ or $dz\wedge d\bar z$. They are related by
$$
dz\wedge d\bar{z}=(d\sigma_1+i d\sigma_2)\wedge(d\sigma_1-id\sigma_2)=-2i
d\sigma_1\wedge d\sigma_2=-2id^2\sigma
$$

The homology group of $\Sigma_g$, $H_1(\Sigma_g)$, is generated by the
homology cycles,
$$
a_1,a_2,\ldots, a_g~~,~~b_1, b_2, \ldots, b_g  
$$
which can be chosen so that they have the standard intersection 
properties
\begin{equation}
a_i\cap a_j=\emptyset~,~b_i\cap
b_j=\emptyset~,~ a_i\cap b_j=\delta_{ij}
\label{homology}
\end{equation}
The first cohomology group $H^1(\Sigma_g)$ is generated by $g$ holomorphic
differentials
$$
\omega_1,\ldots, \omega_g
$$
and their conjugates, the anti-holomorphic differentials.

Linear combinations of these can always be chosen so that they obey
\begin{equation}
\oint_{a_i}\omega_j=\delta_{ij}
~~~,~~~
\oint_{b_i}\omega_j=\Omega_{ij}
\label{orthog}
\end{equation}
where $\Omega_{ij}$ is a $g\times g$ complex, symmetric matrix, called
the period matrix of $\Sigma_g$.   It has positive definite imaginary part.

We shall use the Riemann bilinear relation
\begin{equation}
\int\omega_i\wedge \bar\omega_j=
\sum_{k=1}^g\left(\oint_{a_k}\omega_i\oint_{b_k}\bar\omega_j-
\oint_{b_k}\omega_i\oint_{a_k}\bar\omega_j\right)=
-2i\left(\Omega_2\right)_{ij}
\label{rbr}
\end{equation}

\subsection{String Action}

The string path integral for the vacuum energy is
\begin{eqnarray}
F&=&-
\sum_{g}g_s^{2g-2} \int [dh_gdXd\Psi]
\exp\left[-\frac{1}{4\pi\alpha'}\int_{\Sigma_g} 
d^2\sigma \left( \sqrt{h}h^{\alpha \beta}
\partial_{\alpha}X^\mu\partial_{\beta}X^\mu \right.\right.\cr
&-&\left.\left.2\pi\alpha'iB^E_{\mu\nu}
\epsilon^{\alpha \beta}
\partial_{\alpha}X^\mu\partial_{\beta}X^\nu \right)\right]
\label{partf}
\end{eqnarray}
 The
string coupling constant is $g_s$ and its powers weight the genus,
$g=0,1,\ldots$, of the string's world-sheet.  
For each value of the genus, $g$, $[dh_g]$ is an
integration measure over all metrics of that genus and is normalized
by dividing out the volume of the world-sheet re-parametrization and
Weyl groups.  We will assume that the metrics of both the world-sheet
and the target spacetime have Euclidean signature.

We shall consider the situation in which the target space is 
compactified according to eqs.(\ref{lcc},\ref{x0c}).
Compactification is implemented by including the possible wrappings of
the string worldsheet on the compact dimensions.  These form distinct
topological sectors in the path integration in (\ref{partf}).  In the
wrapping sectors, the bosonic coordinates of the string should have a
multi-valued part which changes by $\beta\cdot$integer or
$(i)\sqrt{2}R\cdot$integer as it is transported along a homology cycle.  
This is accomplished by adding a multi-valued classical piece to the embedding
coordinate of the string, 
\be
X=X_{\rm cl}+X_q
\label{xclxq}
\ee
where the quantum part $X_q$ 
are the variables that are to be integrated in the path integral.  The 
classical part contains the topological part of the configuration.
$X_{\rm cl}$ must be harmonic, so that it is a solution of the equation of
motion, $\partial\bar\partial X=0$.  When $X_{\rm cl}$ is 
transported around  
a homology cycle $X_{\rm cl}$ should change by an integer multiples of the 
compactification circumferences.  The integers are the number of times the
homology cycle of the worldsheet wraps the compactified spacetime dimension. 

It is convenient to work with the differential
$dX_{\rm cl}$ which we assume is single-valued.
It can be decomposed in complex differentials as
\begin{equation}
dX_{\rm cl}=\partial X_{\rm cl}+\bar\partial X_{\rm cl}
\label{exd}
\end{equation}
where
$$
\partial X_{\rm cl}=\partial_zX_{\rm cl} dz~,~~~~~~~~~
\bar\partial X_{\rm cl}=\partial_{\bar z}X_{\rm cl} d\bar z
$$
Since $X_{\rm cl}$ is harmonic, these can be expanded in the
basis for holomorphic and anti-holomorphic differentials,
$\omega_i$ and  $\bar\omega_i$, as
\begin{equation}
\partial X_{\rm cl}=\sum_{i=1}^g \lambda_i\omega_i~,~~~~~~~
\bar\partial X_{\rm cl}=\sum_{i=1}^g\bar\lambda_i\bar\omega_i
\end{equation}
with suitable complex coefficients $\lambda_i$. $\bar\lambda$ is not 
necessarily the complex conjugate of $\lambda$ when there are complex 
identifications of the embedding variables.
In particular, if we choose
\begin{equation}
dX^0_{\rm cl}=\sum_{i=1}^{g}
\left( \lambda_i\omega_i+\bar\lambda_i\bar\omega_i\right)
~~,~~
dX^{1}_{\rm cl}=\sum_{i=1}^{g}\left( 
\gamma_i\omega_i+\bar\gamma_i\bar\omega_i\right)
\label{coordhg}
\end{equation}
Then, we require
\begin{equation}
\oint_{a_i}dX^0_{\rm cl}=\beta m_i+\sqrt{2}\pi Ri p_i
~~~,~~~
\oint_{b_i}dX^0_{\rm cl}=\beta n_i+\sqrt{2}\pi Ri q_i
\end{equation}
\begin{equation}
\oint_{a_i}dX^{1}_{\rm cl}=\sqrt{2}\pi R p_i
~~~,~~~
\oint_{b_i}dX^{1}_{\rm cl}=\sqrt{2}\pi R q_i
\end{equation}
With (\ref{orthog}) and (\ref{rbr}) we use 
these equations to solve for the constants in
(\ref{coordhg}). We find
\begin{eqnarray}
\lambda &=&
\frac{i}{2}\Omega_2^{-1}\left[ \bar\Omega (\beta m+i\sqrt{2}\pi R p)-
(\beta n +i\sqrt{2}\pi R q)
\right]\cr
\bar\lambda &=&-\frac{i}{2}\Omega_2^{-1}\left[\Omega 
(\beta m+i\sqrt{2}\pi R p)
-(\beta n +i\sqrt{2}\pi R q)
\right]\cr
\gamma &=&\frac{i\sqrt{2}\pi R}{2}\Omega_2^{-1}\left(\bar\Omega p- q
\right)\cr
\bar\gamma &=&-\frac{i\sqrt{2}\pi R}{2}\Omega_2^{-1}\left(\Omega  p- q
\right)
\label{llgg}
\end{eqnarray}
Where we have used an obvious vector notation for multiplying the period
matrix and integer-valued vectors $p_i$ and $q_i$, etc.

The string action for the classical configurations
\begin{eqnarray}
S_{\rm cl}&=&\frac{1}{4\pi\alpha'}\int_{\Sigma_g} d^2\sigma
 \left(\partial_{\alpha} X_{\rm cl}^\mu\partial_{\alpha} X_{\rm cl}^\mu-
2\pi\alpha'iB^E_{\mu\nu}
\epsilon^{\alpha \beta}
\partial_{\alpha}X_{\rm cl}^\mu\partial_{\beta}X_{\rm cl}^\nu\right)\cr
&=&
\frac{i}{2\pi\alpha'}\int_{\Sigma_g} 
 \left(\partial X_{\rm cl}^\mu\wedge\bar
\partial X_{\rm cl}^\mu-
2\pi \alpha'B^E_{\mu\nu}
\partial X_{\rm cl}^\mu\wedge\bar\partial X_{\rm cl}^\nu\right)
\end{eqnarray}
using (\ref{rbr}) and (\ref{coordhg}), becomes
\begin{equation}
S_{\rm cl}=\frac{1}{\pi\alpha'}\left[\lambda\Omega_2\bar\lambda+
\gamma\Omega_2\bar\gamma+
2\pi \alpha' B^E
\left(\lambda\Omega_2\bar\gamma-\gamma\Omega_2\bar\lambda\right)\right]
\end{equation}
where, as usual, $B^E=B^E_{01}$.
The part of the string action which contains the winding integers,
can then be obtained by the solutions (\ref{llgg}) and reads
\begin{eqnarray}
S &=\frac{\beta^2}{4\pi\alpha'}\left(
m\Omega^{\dagger}-n\right)\Omega_2^{-1} \left( \Omega m-n\right)+ 2\pi
i \frac{ \sqrt{2}\beta R} {4\pi\alpha'}\frac{1}{2}\left[\left(
p\Omega^{\dagger}-q\right)\Omega_2^{-1} \left( \Omega m-n\right)
\right. \nonumber \\ 
&+\left.\left(
m\Omega^{\dagger}-n\right)\Omega_2^{-1} \left( \Omega
p-q\right)-4\pi\alpha' B^E(mq-np)\right]
\end{eqnarray}
As expected the $B$-part of the action does not depend on the period 
matrix and it is a purely topological term.
Note that the integers $p_i$ and $q_i$ appear linearly in a purely
imaginary term in the action.  When
the action is exponentiated and summed over $p_i$ and $q_i$, the
result will be periodic Dirac delta functions.  It can be shown that
these delta functions impose a linear constraint on the period matrix
of the world-sheet.  Thus, with the appropriate Jacobian factor, the
net effect is to insert into the path integral measure the following
expression,
\begin{eqnarray}
&&\sum_{mnks}e^{-\frac{\beta^2}{4\pi\alpha'}\left(
m\Omega^{\dagger}-n\right)\Omega_2^{-1} \left(\Omega m-n\right)}
\left(\frac{\nu^2}{1+(2\pi\alpha'B^E)^2}\right)^g \left| 
\det\Omega_2\right| \cr
&&\prod_{j=1}^g
\delta\left(\sum_{i=1}^g\left( k_i+\tau_0 m_i
\right)\Omega_{ij}-\left(s_j+\tau_0
n_j \right) \right)\quad
\label{mod}
\end{eqnarray}
where $\tau_0$ is defined in (\ref{tauzero}) and 
$ \nu=4\pi\alpha'/\sqrt{2}\beta R$ as in (\ref{nu}). 
Consequently, the integration over metrics in the string path
integral is restricted to those for which the period matrix obeys
the constraint
\begin{equation}
\sum_{i=1}^g\left(k_i+\tau_0
m_i\right)\Omega_{ij}-\left(
s_j+\tau_0  n_j\right)=0
\label{const}
\end{equation}
for all combinations of the $4g$ integers $m_i,n_i,k_i,s_i$ such 
that $\Omega_{ij}$ is in a fundamental domain ${\cal F}$ of the Riemann surface.
For genus 1 this equation provides the discrete Teichm\"uller parameter
defined in (\ref{dtpm}).
 
The Riemann surfaces with the constraint (\ref{const}) have been classified
in ref.\cite{Grignani:2000zm}. They are the branched coverings of the torus
with Teichm\"uller parameter 
$\tau_0$.

\section{Superstring}

We shall now consider the DLCQ of a closed  superstring  
in ten dimensions in the presence of a $B$-field.  
In this section we shall 
derive, following Atick and Witten~\cite{Atick:1988si},
the expression of the free energy
using the path integral for the superstring in the
Neveu-Schwarz-Ramond ($NSR$) formalism. By means of modular 
invariance \cite{O'Brien:1987pn}
\cite{McClain:1987id}, we shall then obtain the form of the 
superstring free energy that has to be compared with the one of the
matrix string theory.

The strategy we shall adopt is the one described in section 3,
namely we will use a compactified light-like direction $X^+$
and a real Euclidean $B^E$ field and only at the end of the 
calculation we shall analytically continue the result to Minkowski
space $B^E\to -iB$.
 
The superstring covariant action in the $NSR$ formalism in the  
presence of an antisymmetric tensor $B_{\mu\nu}$ is given by 
\begin{eqnarray} 
S&=& 
\frac{1}{4\pi\alpha'}\int d^2 {\sigma}(\sqrt{g}g^{\alpha\beta} 
\partial_{\alpha} X^\mu \partial_{\beta} X_\mu -i 2\pi\alpha' B^E_{\mu\nu} 
\epsilon^{\alpha\beta}\partial_{\alpha} X^\mu \partial_{\beta} X^\nu \cr 
&-&2i\alpha ' e \bar\psi^\mu \rho^a\alpha e_a^\alpha\partial_\alpha\psi_\mu  
-\pi  (2\alpha ')^2 B^E_{\mu\nu}\epsilon^{\alpha\beta} 
\bar\psi^\mu \rho_a e_\alpha^a  
\partial_\beta\psi^\nu )  
\label{nsraction} 
\end{eqnarray} 
Here $\psi^{\mu}$ is a Majorana spinor,  
$g_{\alpha\beta}$ is the world-sheet metric, 
$e_\alpha^a$ the corresponding $zweibein$ and $e$ its determinant. 
At one-loop, $i.e.$ on a torus, we shall take for $g_{\alpha\beta}$  
the metric given in (\ref{wsmetric}). 
To provide the free energy, this action has to be  
inserted in the string path integral,
the metrics of both the world sheet and the  
target spacetime  having Euclidean signature.

For a constant $B$-field, the contribution  
of the $B$-field term to the world-sheet Lagrangian density is  
proportional to a total derivative. Since the boundary conditions on the 
fermions are periodic (Ramond sector) or anti-periodic (Neveu-Schwarz sector) 
the coupling of $B$ to the $\psi^\mu$ in (\ref{nsraction}) actually vanishes. 
Thus the only $B$-dependent contribution comes from the bosonic part 
of the action. 
 
The boundary conditions we shall consider on the bosonic coordinates  
are given by eqs.(\ref{boundcs}) with $R_s=0$. The compactification of 
the coordinate  $X^0$, with the appropriate modification 
of the GSO projection to make  
space-time fermions anti-periodic, introduces the temperature, so that 
the path integral with the action (\ref{nsraction}) computes the  
thermodynamic free energy. 
 
The part of the string action 
which contains the winding integers, $S_{w.m.}$, reads 
\begin{eqnarray} 
S_{w.m.}(m,n,p,q) 
&=&\frac{\beta^2}{4\pi \alpha '\tau_2}\left(m\bar\tau -n\right) 
\left(\tau m-n\right)+2\pi i\frac{\sqrt 2 \beta R}{4\pi \alpha '\tau_2} 
\left[nq+|\tau |^2mp \right. \cr 
&-&\left. \tau_1(mq+np)\right]-2\pi i\sqrt 2 \beta R\frac{B^E}{2}(mq-np) 
\end{eqnarray} 
Note that the integers $p$ and $q$ appear linearly in a purely imaginary term. 
Furthermore, since they will come from the compactification of the light-cone  
direction, this is the only place that they will appear in the string path  
integral. Unlike $m$ and $n$ which appear in the weights of  
the sum over spin structures, the compactification 
of the light-cone direction does not modify the GSO projection. 
Thus, when the action is exponentiated and summed over $p$ and $q$ the result 
will be, as in the bosonic case, periodic delta functions. 
 
The type II superstring free energy in the Neveu-Schwarz-Ramond formulation 
of superstring in a $B$-field background can then be obtained
as in \cite{Atick:1988si}. It reads   
\begin{eqnarray} 
\frac{F}{V}&=&-\frac{1}{4}\left(\frac{1}{4\pi^2\alpha'}\right)^5\int_{\cal F} 
\frac{d^2\tau}{\tau^6_2}\left|\frac{1}{\eta(\tau)}\right|^{24} 
\sum_{n,m,p,q=-\infty}^{+\infty} 
e^{-S_{w.m.}(m,n,p,q)}\cr 
&&\left[\left(\theta_2^4\bar\theta_2^4+\theta_3^4\bar\theta_3^4+ 
\theta_4^4\bar\theta_4^4\right)(0,\tau)+e^{i\pi(n+m)} 
\left(\theta_2^4\bar\theta_4^4+\theta_4^4\bar\theta_2^4\right)(0,\tau) \right.\cr 
&& \left.-e^{i\pi n} 
\left(\theta_2^4\bar\theta_3^4+\theta_3^4\bar\theta_2^4\right)(0,\tau) 
-e^{i\pi m} 
\left(\theta_3^4\bar\theta_4^4+\theta_4^4\bar\theta_3^4\right)(0,\tau)\right] 
\label{fensr}
\end{eqnarray} 
where the Teichm\"uller parameters are integrated over the region ${\cal F}$, 
which is the fundamental domain of the torus. 
Characterizing these wrappings 
by a single integer~\cite{O'Brien:1987pn}
\cite{McClain:1987id}, the integration domain for 
the Teichm\"uller parameters 
is expanded from ${\cal F}$ to the region 
${\cal S}$ defined in (\ref{cals}).
 
If set $m=0$ and integrate 
over the region ${\cal S}$ in (\ref{fensr}) we get \cite{Grignani:1999sp}
\begin{eqnarray}
\frac{F}{V}=-\frac{1}{4}\left(\frac{1}{4\pi^2\alpha'}\right)^5
\int^{1/2}_{-1/2} d\tau_1\int_0^\infty
\frac{d\tau_2}{\tau^5_2}\left|\frac{1}{\eta(\tau)}\right|^{24}
\sum_{n,k,s=-\infty}^{+\infty}\frac{1}{k^2}
e^{-\beta^2 n^2/(4\pi\alpha'\tau_2)}\cr C(n,\tau) 
\delta \left(\tau_1
-\frac{2\pi \alpha 'B^E n}{\nu k}-\frac{s}{k}\right) 
\delta \left(\tau_2-\frac{n}{\nu k} 
\right) 
\label{Faw}
\end{eqnarray}
where $C(n,\tau)$ is the following combination of thetas 
\begin{eqnarray}
C(n,\tau)&=&\left[\left(\theta_2^4\bar\theta_2^4+\theta_3^4\bar\theta_3^4+
\theta_4^4\bar\theta_4^4-
\theta_3^4\bar\theta_4^4-\theta_4^4\bar\theta_3^4\right)(0,\tau)\right.\cr
&+& \left.e^{i\pi n}
\left(\theta_2^4\bar\theta_4^4+\theta_4^4\bar\theta_2^4
-\theta_2^4\bar\theta_3^4-\theta_3^4\bar\theta_2^4\right)(0,\tau)\right]
\end{eqnarray}
$C(n,\tau)$ can be rearranged according to
\begin{equation}
C(n,\tau)=\left(\theta_3^4(0,\tau)-\theta_4^4(0,\tau)-
e^{i\pi n}\theta_2^4(0,\tau)\right)\left(\bar\theta_3^4(0,\tau)-
\bar\theta_4^4(0,\tau)-e^{i\pi n}\bar\theta_2^4(0,\tau)\right) 
\end{equation}
Using Jacobi's identity (\ref{abstruse}) 
$C(n,\tau)$ becomes
\begin{equation}
C(n,\tau)=2\theta_2^4\bar\theta_2^4(0,\tau)(1-e^{i\pi n})=
2\left|\theta_2(0,\tau)\right|^8(1-e^{i\pi n})
\end{equation}
The free energy (\ref{Faw}) then reads
\begin{eqnarray}
\frac{F}{V}=&-&\left(\frac{1}{4\pi^2\alpha'}\right)^5
\int^{1/2}_{-1/2} d\tau_1\int_0^\infty
\frac{d\tau_2}{\tau^5_2}\left|\frac{1}{\eta(\tau)}\right|^{24}
2\sum_{\stackrel{n=1}{n\ odd}}^{+\infty}
\sum_{k=1}^{+\infty}\sum_{s=0}^{k-1}\frac{1}{k^2}
e^{-\frac{\beta^2 n^2}{4\pi\alpha'\tau_2}}\cr
& &
\left|\theta_2(0,\tau)\right|^8
\delta \left(\tau_1 
+\frac{2\pi \alpha 'B^E n}{\nu k}-\frac{s}{k}\right)
\delta \left(\tau_2-\frac{n}{\nu k} 
\right) ~~ 
\label{Faw1}
\end{eqnarray}

The product representation of $\theta_2(0,\tau)$ in (\ref{triple})
implies that
\begin{equation}
\left|\frac{1}{\eta(\tau)}\right|^{24} \left|\theta_2(0,\tau)\right|^8
=2^8\prod_{n=1}^{\infty}
\left|\frac{1+e^{2\pi i n\tau}}{1-e^{2\pi i n\tau}}\right|^{16}
\equiv 2^8 \left|\theta_4\left(0,2\tau\right)\right|^{-16}
 \end{equation}

When the $B^E$ field is  
continued to Minkowski space $B^E\to - i B$. 
the free energy for  
the type II superstring becomes 
\begin{equation} 
\frac{F}{V}=-\sum_{\stackrel {n=1} {n ~\rm odd} }^\infty 
\sum_{k=1}^\infty 
\sum_{s=0}^{k-1}\frac{1}{k^2}\left( \frac{1}{4\pi^2\alpha'\hat{\tau}_2} 
\right)^5 
2^9 \left(\theta_4\left(0,2\tau_- \right)\right)^{-8} 
\left(\bar{\theta}_4\left(0,2\tau_+  \right)\right)^{-8} 
e^{-n^2\beta^2/4\pi\alpha'\tau_2} 
\label{supfe1} 
\end{equation} 
where the Teichm\"uller parameters $\tau_+$ and $\tau_-$ are defined in  
(\ref{tauplus}) and (\ref{tauminus}). 
 
This is the free energy of the DLCQ type II superstring in a constant  
$B$-field which, in the next sections, will be compared with the one derived
for the matrix string theory.

\section{The thermodynamic partition function of the matrix string 
theory in a B field}

In this section we shall first find the Matrix string theory 
action in the presence of a background $B$-field whose only non-zero
component is $B_{01}$. Then we will consider the resulting Matrix string
theory in the limit which should coincide with free type IIA strings.
We shall show that, when the thermodynamic 
partition function of Matrix string theory in this limit 
is computed, it coincides 
with the one obtained in the previous section for the DLCQ type II  
superstring.  

The matrix model of $M$-theory is the effective field theory of the low 
energy interactions of a gas of D0-branes propagating on 10-dimensional
Minkowski space.  When one spatial direction is compactified, the matrix
model which describes the resulting theory is 1+1-dimensional maximally
supersymmetric Yang-Mills theory.  This theory contains $T$-duality 
in a manifest
way - it can either describe D0-branes or their $T$-dual, 
D1-branes which wrap
the compact direction where the compactification radius $R_1$ 
should be replaced
by the dual radius $\alpha'/R_1$.  To find the coupling of a background 
$B$-field to
D0-branes on Minkowski space with a compact dimension, it is convenient 
to begin with
the $T$-dual of that situation which is wrapped D1-branes, no 
$B$-field, and space-time
with the metric (\ref{metric1}).
 
The 1+1-dimensional Yang-Mills theory which describes this system is gotten by
starting with ten-dimensional $U(N)$ super Yang-Mills theory 
in the background metric (\ref{metric1}) \cite{Taylor:1999gq}.  
In Euclidean space time,
the ten dimensional action is 
\be
S=\frac{1}{2}\int d^{10}\xi{\rm Tr}\sqrt{G}\left(
F_{\mu\nu}G^{\mu\rho}G^{\nu\sigma}F_{\rho\sigma}
-\frac{i}{\pi}{\rm Tr}\bar\psi\gamma^a V^\mu_a D_\mu\psi\right)
\label{sym}
\ee
where the fields $A_\mu$ and $\psi$ are both in the 
adjoint representation of $U(N)$.
The covariant derivative is given by
$$
D_\mu=\partial_\mu - i g_{{\rm YM}}\left[ A_\mu,\dots \right]
$$
$G_{\mu\nu}$ is a metric and
$V_\mu^a$ the corresponding $zehnbein$.
In Euclidean space, in terms of the Euclidean $B^E$-field the metric reads
\be
G^{\mu\nu}= \left( \matrix{
1+(2\pi\alpha'B^E)^2 & -2\pi\alpha'B^E & 0 & ... \cr -2\pi\alpha' B^E & 1 & 0
& ... \cr 0 & 0 & 1 & ... \cr ....  & ...  & ...& ...\cr } \right) 
\label{metric}
\ee
Ten-dimensional Yang-Mills theory is reduced to 2-dimensional Yang-Mills 
theory by
assuming that the fields are independent on the coordinates $2,\dots, 9$.
In 1+1 dimensions
there is a gauge field $A_\alpha$, $\alpha=1,2$, 8 adjoint scalars 
$X^i,\ i=2,\dots 9$ and a 16-component Majorana spinor which can by arranged
into 8 two-dimensional Majorana spinors.  The action (\ref{sym}) becomes
\bea
S&= &\frac{1}{2} \int d^2\sigma \left[\sqrt{g} g^{\alpha\beta}D_\alpha X^i
D_\beta X^i-g^2_{\rm YM}\sum_{i<j}\left[X^i,X^j\right]^2
+\sqrt{g}F_{\alpha\beta}g^{\alpha\gamma}g^{\beta\delta}
F_{\gamma\delta} \right.\cr
&-&\left. \frac{i}{\pi} \left(\sqrt{g}\psi^T\gamma^a e^\alpha_a D_\alpha\psi+
g_{\hbox{\tiny{\rm YM}}}\psi^T\gamma_i[X^i,\psi]\right)
\right]
\label{d1a}
\eea
where $g_{\hbox{\tiny{\rm YM}}}$ is the ten-dimensional 
Yang-Mills coupling constant,
$g^{\alpha\beta}$ is the $(0,1)$ part of the metric (\ref{metric})
and $e^\alpha_a$ the corresponding $zweibein$.
This is the action describing the low energy dynamics of $N$ coincident
$D1$-brane in the background metric $g_{\alpha\beta}$.
It can be easily seen that the linear terms in $B$ in (\ref{d1a})
are exactly those found in \cite{Schiappa:2000dv,Taylor:2000pr}.
When the direction 1 is compactified, 
using the Buscher rules for $T$-duality, (\ref{d1a}) is equivalent to 
$D0$-branes on a space with the same compactified direction, 
the Euclidean metric and a non-zero $B$-field in the 
$\left(0,1\right)$ direction.

To arrive at the matrix string point of view~\cite{Dijkgraaf:1997vv}, 
(which we shall refer to as DVV) we have to take into account that, 
conventionally, the matrix theory 
is related to $M$-theory via the compactification of the 11-th
direction on a circle of radius $R_{11}$. 
We then re-scale the field variables, the coupling 
constant and  the coordinates in (\ref{d1a}) in such a way 
to introduce the compactification 
radius $R_{11}$, the  string tension $1/2\pi\alpha'$ and
coupling constant $g_s$. The necessary re-scalings are
\bea
&X^i&\to \left(\frac{1}{2\pi\alpha'}\right)^{1/2}X^i~,~~~~
\psi\to \left(\frac{R_{11}^2}{2\pi\alpha'}\right)^{1/4}\psi\cr
&A^1&\to \left(\frac{R^2_{11}}{2\pi\alpha'}\right)^{1/2}A^1~,~~~~
A^2\to\left(2\pi\alpha' \right)^{1/2}A^2~,~~~~
g_{\hbox{\tiny{\rm YM}}}\to \frac{R_{11}}{g_s\sqrt{\alpha'}}\cr
&\sigma^1&\to
\left(\frac{2\pi\alpha'}{R^2_{11}}\right)^{1/2}\sigma^1~,~~~~\sigma^2\to
\left(\frac{1}{2\pi\alpha'}\right)^{1/2}\sigma^2
\eea
The action (\ref{d1a}) then becomes
\bea
S_{\rm DVV}&=&\int d^2 \sigma\left\{\frac{1}{2 R_{11}}\left[\left(D_2 X^i- 
R_{11} B^E D_1X^i\right)^2 + \left(\frac{R_{11}}{2\pi\alpha'} D_1X^i\right)^2 
+{g^2_s\alpha'}F^2_{\mu\nu}\right.\right.\cr
&-&\left.\left.
\frac{R^2_{11}}{4\pi^2\alpha'^3 g^2_s}\sum_{i<j}[X^i,X^j]^2\right]
-\frac{i R_{11}}{2\pi}\psi^T\left(\frac{1}{R_{11}}D_2\psi\right.\right.\cr
&+&\left.\left.\frac{1}{2\pi\alpha'}\gamma D_1\psi -B^E D_1\psi -
\frac{i}{2\pi\alpha'^{3/2}g_s}\gamma_i[X^i,\psi]\right)\right\}
\label{sdvv}
\eea
The two dimensional world-sheet here is a cylinder with compact spatial coordinate
$\sigma_1\in[0,1)$ and Euclidean time $\sigma^2\in (-\infty,\infty)$.  Later, when
we consider the Euclidean action in the partition function, Euclidean time will also
be compact.
This is the action that, according to the DVV construction
~\cite{Dijkgraaf:1997vv}, should describe 
the non-perturbative particle spectrum of the type IIA string in a $B$-field
in terms of the states that can be made up from infinitely many $D0$-branes.
The compactification  on the circle $S^1$ along the 1-direction and the 
reinterpretation of the $X^1$ as a covariant derivative $D_1$ with 
a gauge field defined on $S^1$, amounts to applying a $T$-duality 
transformation along the $S^1$ direction, thereby turning the $D0$-branes 
into $D1$-branes. This is the reason why this action can be obtained 
from the $D1$-brane action (\ref{d1a}).

The Hamiltonian corresponding to the action (\ref{sdvv}) and that will 
be used in the calculation of the thermodynamic partition function, reads
\bea
H&=&\frac{R_{11}}{2}\int_0^1d\sigma_1\tr\left[\Pi_i^2+
2iB^E \Pi^i D_1 X^i+\frac{1}{4\pi^2
\alpha'^2}(D_{1} X^i)^2 -\frac{1}{4\pi^2 \alpha'^3 g_s^2}
\sum_{i<j}\left[X^i,X^j\right]^2 \right.\cr&&\left.
+\frac{E^2}{g_s^2\alpha'} +  
\frac{i }{\pi}B^E \psi^T \cdot D_{1}\psi-
\frac{i }{2\pi^2\alpha'}\psi^T \gamma\cdot D_{1}\psi-
\frac{1}{2\pi^2\alpha'^{3/2} g_s}\psi^T\gamma_i\left[X^i,\psi\right]\right]
~~~~~~~~
\label{HDVV}
\eea
The mass-dimensions of $H$ and of the fields are 
$$
[H]=M\quad,\qquad [X^a]=M^{-1}\quad,\qquad [\psi]=M^0\quad,
\qquad [A^{1,2}]=M^0.
$$
$R_{11}$ is the radius of the dimension that must be compactified 
in order to obtain the matrix description of $M$-theory - in its original 
form the Matrix model describes $M$-theory in a reference frame which has 
infinite momentum in the 11'th direction.   In the more sophisticated 
proposal of ref. \cite{Susskind:1997cw},  
this compactified direction is the light-cone 
direction $X^+$.  It was shown by Seiberg \cite{Seiberg:1997ad} that, 
with certain assumptions, 
a boost to the frame with infinite momentum of a theory compactified on
a small circle $R_{11}$ is equivalent to one with the light-cone compactified.
For a more recent discussion of these issues see 
\cite{Bilal:1998ys,Bilal:1998vq,Hyun:1999iq}.
Here, since we are actually describing DLCQ $M$-theory, 
we shall make the replacement
$$
R_{11}~~\longrightarrow~~R
$$
The hypothesis is that this model describes DLCQ $M$-theory with one
dimension compactified (to describe the DLCQ IIA superstring).
The canonical momenta which appear in (\ref{HDVV}) are normalized so
that they have the conventional canonical commutators,
\begin{eqnarray}
\left[ X^i_{ab}(\sigma), \Pi^j_{cd}(\sigma')\right]&=&
i\delta_{ad}\delta_{bc}\delta^{ij}\delta(\sigma-\sigma')  \cr
\left[ A_{ab}(\sigma), E_{cd}(\sigma')\right]&=&
i\delta_{ad}\delta_{bc}\delta(\sigma-\sigma')  \cr
\left\{ \psi_{ab}(\sigma), \psi_{cd}(\sigma')\right\}&=&\frac{1}{2\pi}
\delta_{ad}\delta_{bc}\delta(\sigma-\sigma')
\end{eqnarray}
Note that $g_s^2\alpha'$ plays the role of the inverse square of the
Yang-Mills coupling constant in 1+1 dimension: 
$g_s^2\alpha'=g^{-2}_{\rm YM}$

\subsection{Matrix theory in a $B$-field}

In this subsection, we will find the matrix model which describes $M$-theory 
in 
a $B$-field on an 11-dimensional space whose only compactification is the null
direction $X^+$.  To do this, 
we shall follow an inverse DVV procedure~\cite{Dijkgraaf:1997vv}.
This result will be achieved by first  decompactifying the $T$-dual circle 
of radius $R_1$ and then interchanging the role of the $11^{th}$
and the $1^{st}$ directions. 

To this purpose we first rewrite
the Hamiltonian (\ref{HDVV})
in terms of the 11-dimensional Plank length $l_p$, the radius of the 
compactified direction, $R_1$, and the string coupling constant
$g_s$. The relation between these constants and the string scale 
$\sqrt{\alpha'}$ is defined as
\be
g_s\sqrt{\alpha'}=R_1~,~~~~~~~~~~~~~~g_s^{1/3}\sqrt{\alpha'}=l_p
\ee
This leads to the Hamiltonian
\bea
H&=&\frac{R_{11}}{2}\int_0^1d\sigma_1\tr\left[\Pi_i^2+\frac{R_1^2}{4\pi^2
l_p^6}(D_{1} X^i)^2 -\frac{1}{4\pi^2 l_p^6}\sum_{i<j}\left[X^i,X^j\right]^2
+2i B^E D_1 X^i\Pi^i\right.\cr
&+&\left.\frac{E^2}{R_1^2} +\frac{i}{\pi}B^E \psi^T D_1\psi -  
\frac{1}{2\pi^2 l_p^3}\psi^T\gamma_i\left[X^i,\psi\right]-
\frac{i R_1}{2\pi^2 l_p^3}\psi^T \gamma\cdot D_{1}\psi\right]
\label{hdvvlp}
\eea
where $i=1,\dots,8$ now labels the transverse coordinates.
We now decompactify the $1^{st}$ direction, so that, after the usual 
$T$-duality,
we can identify the covariant derivative $D_1$ with the $X^1$ 
coordinate, according to
\be
i R_1 D_1\longrightarrow X^1
\ee
With this procedure the Hamiltonian (\ref{hdvvlp}) becomes
\bea
H=\frac{R_{11}}{2}{\rm Tr}\left[\Pi_a^2+2\frac{B^E\alpha'}{l_p^3}[X^1,X^a] 
\Pi^a -\frac{1}{4\pi^2l_p^6}[X^a,X^b]^2\right.\cr\left.
-\frac{1}{2\pi^2l_p^3}\psi^T\gamma_a[X^a,\psi]+
\frac{B^E\alpha'}{\pi l_p^3}\psi^T [X^1,\psi]\right]
\eea
where $a=1,\dots,9$.
We now interchange the $11^{th}$ and the $1^{st}$ directions
by introducing eleven-dimensional Planck  units:
$R_{11}=g\sqrt{\alpha'}$ and $l_p^2=\alpha'g^{2/3}$.
The 1-11 flip, has been shown to correspond to a chain of $T-S-T$ duality in
\cite{Schiappa:2000dv}.
With this substitution we obtain the final result for the Hamiltonian
of the Matrix model in a constant $B$-field in the $(0,1)$ direction
\bea
H=\frac{1}{2g\sqrt{\alpha'}}{\rm Tr}\left[g^2\alpha'\Pi_a^2+
2B^E g\sqrt{\alpha'}\Pi^a[X^1,X^a] 
 -\frac{1}{(2\pi\alpha')^2}[X^a,X^b]^2\right]\cr
-\frac{1}{2\pi}{\rm Tr}\left[\frac{1}{2\pi\alpha'}
\psi^T\gamma_a[X^a,\psi]-
2B^E \psi^T [X^1,\psi]\right]
\eea
When the $\sigma_2$ coordinate is identified with the Euclidean time
$\tau$ the Matrix model action then reads 
\begin{eqnarray}
S=\int dt\left\{ \frac{1}{2g\sqrt{\alpha'}}{\rm Tr}
\left[\left(D_{\tau}X^a + iB^E
[X^1,X^a]\right)^2-\frac{1}{(2\pi\alpha')^2}[X^a,X^b]^2\right]  
\right. \nonumber \\
\left.
-\frac{i}{2\pi}\left( \psi^TD_{\tau}\psi +2i B^E\psi^T[X^1,\psi]
 -\frac{i}{2\pi\alpha'}\psi^T\gamma_a[X^a,\psi]\right)\right\}
\label{mma}
\end{eqnarray}
Each of the adjoint scalar matrices $X^a$ is a 
hermitian $N\times N$ matrix, where $N$ is the number of $0$-branes
and the relation between the coupling constant and the 
compactification radius of the $11^{th}$ direction 
$g\sqrt{\alpha'}=R_{11}$, has been used. This action describes a stack
of $N$ $D0$-branes in the presence of a $B$-field in the $(0,1)$ direction.

Since (\ref{mma}) is the action for $N$ coincident $D0$-branes, it should 
be possible to derive it also from dimensional reduction of the 
ten-dimensional $U(N)$ super Yang-Mills theory (\ref{sym}).
Namely we can arrive at (\ref{mma}) by a further dimensional reduction of 
(\ref{d1a}) assuming that all fields are independent of the coordinate
1, calling time the coordinate 2 and performing the following re-scalings of 
the fields, coupling constant and coordinate
\bea
&X^a&\to \left(\frac{1}{2\pi\alpha'g_{\hbox{\tiny{\rm YM}}}}
\right)^{1/2}X^a~,~~~~
A^2 \to \left(2\pi\alpha'g_{\hbox{\tiny{\rm YM}}}\right)^{1/2}A^2\cr
&\sigma^2& \to
\left(\frac{1}{2\pi\alpha'g_{\hbox{\tiny{\rm YM}}}}\right)^{1/2}\tau ~,~~~~
g_{\hbox{\tiny{\rm YM}}}\to \frac{g^2}{2\pi}
\eea

\subsection{Matrix string free energy}

Using the  Hamiltonian (\ref{HDVV}) we shall 
now construct the thermodynamic partition function $Z$ of 
matrix string theory in the limit of vanishing string coupling, $g_s\to 0$. 
The prescription discussed in the previous sections
for the treatment of the thermal ensemble for systems in the light-cone 
frame, and the fact that $P^-=N/R$ lead us to write $Z$ as
\be
Z={\rm Tr} e^{-\beta P^0}={\rm Tr} e^{-\beta\left(P^++P^-\right)/\sqrt{2}}~=~
\sum^\infty_{N=0}~e^{-N\beta/\sqrt{2}R}~{\rm Tr}\left\{e^{-\beta H/\sqrt{2}}
\right\}
\label{Z}
\ee
where $P^+=H$ is the matrix string theory Hamiltonian in (\ref{HDVV}).
The trace of $\exp(-\beta H/\sqrt{2})$ is to be taken over gauge invariant
states of the two-dimensional super-Yang-Mills theory.  This trace
has the standard Euclidean field theory 
path integral expression \cite{Gross:1981br}
\begin{equation}
Z[\beta]=\sum_{N=0}^\infty 
\int [dA_\alpha][dX^i][d\psi^a]~\exp\left(-\beta N/\sqrt{2}R-S_E[A,X^i,
\psi]\right)
\label{pf33}
\end{equation}
where $S_E$ is the Euclidean action 
\begin{eqnarray}
S_E&=&\frac{1}{4\pi\alpha'}\int_0^1 d\sigma_1\int_0^1 d\sigma_2\tr\left\{ 
\frac{1+(2\pi\alpha'B^E)^2}{\nu} D_1 X^i D_1 X^i
+\nu D_2 X^i D_2 X^i\right.\cr
&-&\left.4\pi\alpha' B^E D_1X^i D_2X^i+ \nu g_s^2\alpha'
F_{\mu\nu}^2-\frac{1}{\nu\alpha'g_s^2}
\sum_{i<j}\left[ X^i,X^j\right]^2\right.\cr &-&\left. 
i 2\alpha'\left(\psi^T D_2\psi+
\frac{1}{\nu}\psi^T\gamma D_1\psi-
\frac{2\pi\alpha' B^E}{\nu}\psi^T D_1\psi-
\frac{i}{\nu g_s\sqrt{\alpha'}}
\psi^T\gamma^i\left[X^i,\psi\right]\right)
\right\}\nonumber\\
\end{eqnarray}
Here $\nu=4\pi\alpha'/(\sqrt{2}\beta R)$,
we have rescaled the time $\sigma_2$ so that the integration is over
a box of area one, $0\leq\sigma_\mu<1$, and $\beta$ appears
as a factor in various coupling constants.  Boundary conditions in the path
integral are
\begin{eqnarray}
A_\mu(\sigma_1+1,\sigma_2)&=&A_\mu(\sigma_1,\sigma_2) \cr
X^i(\sigma_1+1,\sigma_2)&=&X^i(\sigma_1,\sigma_2) \cr
\psi(\sigma_1+1,\sigma_2)&=&\psi(\sigma_1,\sigma_2) \cr
A_\mu(\sigma_1,\sigma_2+1)&=&A_\mu(\sigma_1,\sigma_2) \cr
X^i(\sigma_1,\sigma_2+1)&=&X^i(\sigma_1,\sigma_2) \cr
\psi(\sigma_1,\sigma_2+1)&=&-\psi(\sigma_1,\sigma_2) 
\label{matbc}
\end{eqnarray}
The anti-periodicity of the fermion field in the Euclidean time 
comes from taking the trace in (\ref{Z}).

The limit $g_s\rightarrow 0$ of this theory was formulated by DVV 
\cite{Dijkgraaf:1997vv}.Their hypothesis is that, 
in this limit, the field configurations which have finite action are
simultaneously diagonalizable matrices.  They can therefore all be
related to diagonal field configurations by a gauge transformation
\begin{eqnarray}
X^i(\sigma)&=&U(\sigma)X^i_D(\sigma)U^{-1}(\sigma) \cr
\psi(\sigma)&=&U(\sigma)\psi_D(\sigma)U^{-1}(\sigma) \cr
A_\mu(\sigma)&=&
iU(\sigma)\left(\partial_\mu-iA^D_\mu(\sigma)\right)U^{-1}(\sigma)
\label{diag}
\end{eqnarray}
$X^i_D$, $\psi_D$ and $A^D_\mu$ are diagonal matrices.  The
fields $X^i$, $\psi$ and $A_\mu$ have the (anti-)periodic 
boundary conditions (\ref{matbc}). The spectrum of (anti-)periodic
matrices must be (anti-)periodic.  However, the individual eigenvalues
need not be.  They can be periodic up to   
permutations (and gauge transformations generated by the Cartan 
sub-algebra), 
\begin{eqnarray}
(A^D_\mu)_a(\sigma_1+1,\sigma_2)&=
&(A^D_\mu)_{P(a)}(\sigma_1,\sigma_2)+2\pi (n_\mu)_a +\partial_\mu\theta_a\cr
(X^i_D)_a(\sigma_1+1,\sigma_2)&=&(X^i_D)_{P(a)}(\sigma_1,\sigma_2) \cr
(\psi_D)_a(\sigma_1+1,\sigma_2)&=&(\psi_D)_{P(a)}(\sigma_1,\sigma_2) \cr
(A^D_\mu)_a(\sigma_1,\sigma_2+1)&=&(A^D_\mu)_{Q(a)}(\sigma_1,\sigma_2) 
+2\pi (m_\mu)_a+\partial_\mu\phi_a\cr
(X^i_D)_a(\sigma_1,\sigma_2+1)&=&(X^i_D)_{Q(a)}(\sigma_1,\sigma_2) \cr
(\psi_D)_a(\sigma_1,\sigma_2+1)&=&-(\psi_D)_{Q(a)}(\sigma_1,\sigma_2) 
\label{diagbc}
\end{eqnarray}
where $a=1,...,N$, $P(a)$ and $Q(a)$ are permutations and $n_\mu,m_\mu$ 
are integers.  Consistency
requires that the two permutations commute,
\begin{equation}
PQ=QP
\end{equation}
To compute the partition function, in the $g_s\rightarrow 0$ limit, 
we should now do the path integral 
(\ref{pf33}) over only the diagonal components of the matrix fields with
the action
\begin{eqnarray}
S_{\rm diag}&=&
\frac{1}{4\pi\alpha'}\int_0^1 d\sigma_1\int_0^1 d\sigma_2\tr\left\{ 
\frac{1+(2\pi\alpha'B^E)^2}{\nu} \partial_1 X_a^i \partial_1 X_a^i
+\nu \partial_2 X_a^i \partial_2 X_a^i\right.\cr
&-&\left.4\pi\alpha' B^E \partial_1X_a^i \partial_2X_a^i+
{g_s^2\alpha' \nu}
\left(\partial_1A_{2a}-\partial_2 A_{1a}\right)^2
\right.\cr &-&\left. 
i 2\alpha'\left(\psi_a^T \partial_2\psi_a+
\frac{1}{\nu}\psi_a^T\gamma \partial_1\psi_a-
\frac{2\pi\alpha' B^E}{\nu}\psi_a^T \partial_1\psi_a\right)\right\}
\label{diagft}
\end{eqnarray}
and with the boundary conditions (\ref{diagbc}). We should then sum over
topologically distinct configurations, characterized by the permutations
$P$ and $Q$ and by the integers $m_\mu,n_\mu$. We shall not discuss the 
validity of the 
assumptions leading to  this starting point in the present  
Paper (see discussions in \cite{Wynter:1997yb,Wynter:1998kd}).  

Here, the eigenvalues of the matrices of the matrix model are functions which will
turn out to live on multiple (un-branched) covers of the torus on which the field
theory in (\ref{diagft}) lives.  The rest o this section is devoted to showing that
the combinatorics of combining the eigenvalues to form these covers can be solved
and that the resulting sum over all topological sectors produces a partition function
which coincides with the one found for the superstring in the previous section.
 
It is easy to show that in the case that we are considering, the gauge field, after
absorbing the gauge transformations in (\ref{diagbc}), decouples as $g_s\to 0$.  This
fact is particular to the genus 1 system that we are specializing to here.  At higher
genera it is claimed that the gauge field part of the partition function produces the
correct factor of $(g_s)^{2g-2}$ which weights the contribution of higher genus covers
~\cite{Bonelli:1999wx}.
 
The partition function that we must compute thus decomposes into topological
sectors which are characterized by the permutations of the eigenvalues,
\be 
Z=\sum^\infty_{N=0}\frac{1}{N!}
e^{-N\beta/\sqrt{2}R}\sum_{P,Q} Z(P,Q)\ ,
\label{Zpq}
\ee  
(where we have suppressed the dependence on the integers in the boundary 
conditions for gauge fields).
Here, for each $N$ we have divided by contribution by the volume of the
Weyl group, $N!$, which reflects the fact that the eigenvalues are defined 
up to a global permutation. (From the $M$-theory point of view, this 
factor gives Boltzmann statistics to the D0-branes.)
The possibility of introducing different weights in the sum over
pairs of commuting permutations has been discussed in~\cite{Billo':1999fb},
\cite{Billo':2001wi}.
Some relationships between these combinatorics and
sigma models on symmetric orbifolds has been discussed in \cite{Fuji:2001fa}
and
\cite{Arutyunov:1999eq,Arutyunov:1998gi}.
We shall show in what follows that the correct genus 1 partition function 
for the type 
IIA string emerges from (\ref{Zpq}) only when all of the pairs $(P,Q)$
have the same weight.

\subsubsection{How permutations determine the world-sheet metric}

The basic combinatorics which obtains the genus 1 (un-branched) covers of the 
torus from the sum over compatible permutations $P$ and $Q$ was discussed
in detail in ~\cite{Grignani:1999sp}.  Here, we shall review the essential 
parts of the argument.

Consider a generic set of fields 
obeying the boundary conditions
\bea
\lambda_a(\sigma_1+1,\sigma_2)=\lambda_{P(a)}(\sigma_1,\sigma_2)\ ,\cr
\lambda_a(\sigma_1,\sigma_2+1)=(-1)^f\lambda_{Q(a)}(\sigma_1,\sigma_2)\ .
\label{pq}
\eea
where $f=0$ for a Bose field and $f=1$ for a fermion.A given permutation
$P$ can be decomposed into cycles - i.e. subgroups of the $N$ elements which 
cycle into each other under repeated application of $P$.  

Consider the subset of the fields  which occur in the cycles
of length $k$ of $P$.  We relabel them so that $P$ acts on them
as
\bea
\lambda_{a^{k\alpha}_n}(\sigma_1+1,\sigma_2)=\lambda_{a^{k\alpha}_{n+1}}
(\sigma_1,\sigma_2)\ ,\cr
\lambda_{a^{k\alpha_n}}(\sigma_1,\sigma_2+1)=(-1)^{f}
\lambda_{a^{k\pi(\alpha)}_{n+s}}(\sigma_1,\sigma_2)\ .
\label{pq1}
\eea
Here $\alpha$ is a label that runs over all of the different cycles of $P$
which have length $k$.  $k$ runs over all lengths of cycles of a given
permutation $P$.

For each such cycle, $\alpha$, we fuse these fields together into
a single function which lives on $k$ copies of the torus and is
periodic in the sense that

\bea
\lambda_{a^{k\alpha}}(\sigma_1+k,\sigma_2)=\lambda_{a^{k\alpha}}
(\sigma_1,\sigma_2)\ .
\eea

Then we consider what a commuting permutation $Q$ can do to these 
fields.  Since $Q$ must commute with $P$, there are only two possibilities
for the action of $Q$. First, $Q$ can permute cycles of $P$ which have
equal size - so it can act as a permutation of the set of all cycles of $P$ 
with length $k$ for example $Q:\alpha\to \pi(\alpha)$.  
Second, it can implement a cyclic permutation 
of the elements within a cycle of $P$ which in the notation above is not a 
translation
of $\sigma^1$, $Q:\sigma^1\to\sigma^1+s(k,\alpha)$. Thus, $Q$ has the action

\bea
\lambda_{a^{k\alpha}}(\sigma_1,\sigma_2+1)=(-1)^{f}
\lambda_{a^{k\pi(\alpha)}}(\sigma_1+s(k,\alpha),\sigma_2)\ .
\label{pq2}
\eea

Then, we consider the decomposition of the action of $Q$ into cycles.
Such a cycle of $Q$ permutes a subset (say of length $r$) 
of the $r_k$ length $k$-cycles of $P$
into each other.  For this cycle of cycles, we again fuse together $r$ copies
of the fields
$\lambda_{a^{k\alpha}}(\sigma_1,\sigma_2)$ so that the resulting field
lives of $r_k$ copies of the torus and 
is periodic
\bea
\lambda_{a^{k,r}}(\sigma_1+k,\sigma_2)=\lambda_{a^{k,r}}
(\sigma_1,\sigma_2)\ ,\cr
\lambda_{a^{k,r}}(\sigma_1,\sigma_2+r)=(-1)^{fr}
\lambda_{a^{k,r}}(\sigma_1+s,\sigma_2)\ .
\label{pq3}
\eea  
where $s=\sum_\alpha s(k,\alpha)~{\rm mod}~k$ is the accumulated shift
for the $r$ elements in the cycle of $Q$.

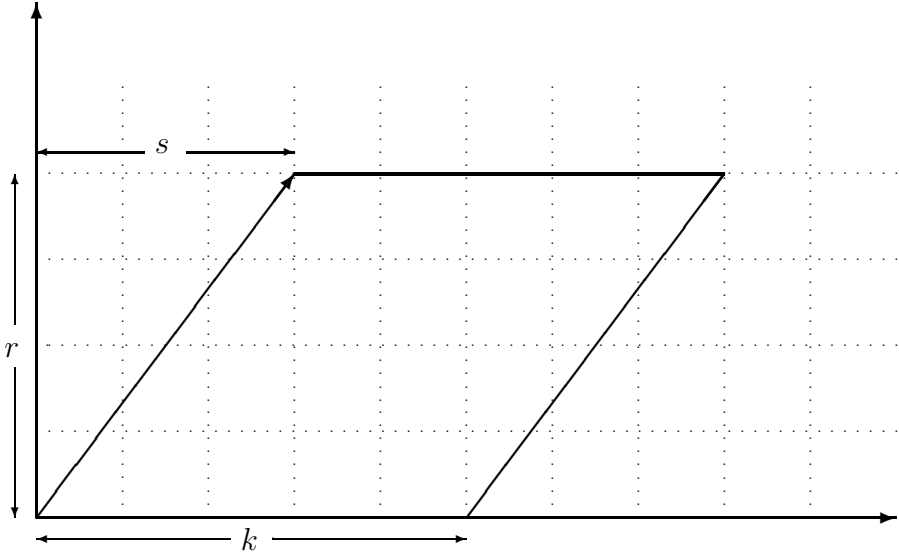
\begin{figure}[htb]
\begin{center}
\setlength{\unitlength}{3552sp}%
\begingroup\makeatletter\ifx\SetFigFont\undefined%
\gdef\SetFigFont#1#2#3#4#5{%
  \reset@font\fontsize{#1}{#2pt}%
  \fontfamily{#3}\fontseries{#4}\fontshape{#5}%
  \selectfont}%
\fi\endgroup%
\begin{picture}(6258,3858)(3376,-4186)
\thinlines
\put(3601,-2161){\makebox(1.8519,12.9630){\SetFigFont{5}{6}{\rmdefault}{\mddefault}{\updefault}.}}
\put(3601,-1636){\makebox(1.8519,12.9630){\SetFigFont{5}{6}{\rmdefault}{\mddefault}{\updefault}.}}
\put(4651,-1411){\vector( 1, 0){750}}
\put(4351,-1411){\vector(-1, 0){750}}
\put(5251,-4111){\makebox(1.8519,12.9630){\SetFigFont{5}{6}{\rmdefault}{\mddefault}{\updefault}.}}
\put(3451,-2611){\vector( 0, 1){1050}}
\put(3451,-2911){\vector( 0,-1){1050}}
\put(4951,-4111){\vector(-1, 0){1350}}
\put(5251,-4111){\vector( 1, 0){1350}}
\thicklines
\put(3601,-3961){\vector( 0, 1){3600}}
\put(3601,-3961){\vector( 1, 0){6000}}
\put(5401,-1561){\line( 1, 0){3000}}
\thinlines
\multiput(3601,-3361)(89.55224,0.00000){68}{\makebox(1.8519,12.9630){\SetFigFont{5}{6}{\rmdefault}{\mddefault}{\updefault}.}}
\multiput(4801,-961)(0.00000,-90.90909){34}{\makebox(1.8519,12.9630){\SetFigFont{5}{6}{\rmdefault}{\mddefault}{\updefault}.}}
\multiput(5401,-961)(0.00000,-90.90909){34}{\makebox(1.8519,12.9630){\SetFigFont{5}{6}{\rmdefault}{\mddefault}{\updefault}.}}
\put(3676,-2761){\makebox(1.8519,12.9630){\SetFigFont{5}{6}{\rmdefault}{\mddefault}{\updefault}.}}
\multiput(6001,-961)(0.00000,-90.90909){34}{\makebox(1.8519,12.9630){\SetFigFont{5}{6}{\rmdefault}{\mddefault}{\updefault}.}}
\put(4426,-2461){\makebox(0,0)[lb]{\smash{\SetFigFont{12}{14.4}{\rmdefault}{\mddefault}{\updefault}}}}
\multiput(6601,-961)(0.00000,-90.90909){34}{\makebox(1.8519,12.9630){\SetFigFont{5}{6}{\rmdefault}{\mddefault}{\updefault}.}}
\multiput(7201,-961)(0.00000,-90.90909){34}{\makebox(1.8519,12.9630){\SetFigFont{5}{6}{\rmdefault}{\mddefault}{\updefault}.}}
\multiput(7801,-961)(0.00000,-90.90909){34}{\makebox(1.8519,12.9630){\SetFigFont{5}{6}{\rmdefault}{\mddefault}{\updefault}.}}
\multiput(8401,-961)(0.00000,-90.90909){34}{\makebox(1.8519,12.9630){\SetFigFont{5}{6}{\rmdefault}{\mddefault}{\updefault}.}}
\multiput(9001,-3961)(0.00000,90.90909){34}{\makebox(1.8519,12.9630){\SetFigFont{5}{6}{\rmdefault}{\mddefault}{\updefault}.}}
\multiput(3601,-2761)(89.55224,0.00000){68}{\makebox(1.8519,12.9630){\SetFigFont{5}{6}{\rmdefault}{\mddefault}{\updefault}.}}
\multiput(3601,-2161)(89.55224,0.00000){68}{\makebox(1.8519,12.9630){\SetFigFont{5}{6}{\rmdefault}{\mddefault}{\updefault}.}}
\multiput(3601,-1561)(89.55224,0.00000){68}{\makebox(1.8519,12.9630){\SetFigFont{5}{6}{\rmdefault}{\mddefault}{\updefault}.}}
\thicklines
\put(6601,-3961){\line( 3, 4){1800}}
\put(3601,-3961){\vector( 3, 4){1800}}
\thinlines
\multiput(4201,-961)(0.00000,-90.90909){34}{\makebox(1.8519,12.9630){\SetFigFont{5}{6}{\rmdefault}{\mddefault}{\updefault}.}}
\put(5026,-4186){\makebox(0,0)[lb]{\smash{\SetFigFont{12}{14.4}{\rmdefault}{\mddefault}{\updefault}$k$}}}
\put(3376,-2836){\makebox(0,0)[lb]{\smash{\SetFigFont{12}{14.4}{\rmdefault}{\mddefault}{\updefault}$r$}}}
\put(4426,-1411){\makebox(0,0)[lb]{\smash{\SetFigFont{12}{14.4}{\rmdefault}{\mddefault}{\updefault}$s$}}}
\end{picture}
\end{center}
\caption{Connected covering torus with $r=4$, $k=5$ and $s=3$.}
\label{fig1}
\end{figure}

The space on which coordinates which are arguments of the field in (\ref{pq3})
take values is the torus depicted in fig. 1.  The contribution to the path
integral of this set of fields is denoted by
\bea
Z(r,k,s)&=&
\int[dX^i][d\psi] \exp\left\{-\frac{1}{2}
\int_0^r d\sigma_2\int_{\frac{s}{r}\sigma_2}^{k+\frac{s}{r}\sigma_2}
d\sigma_1 \right.\cr& &\left.
\left[ \frac{\beta R(1+(2\pi\alpha'B^E)^2)}{\sqrt{2}(2\pi\alpha')^2}
(\partial_1 X^i)^2-
B^E\partial_2X^i\partial_1X^i+
\frac{\sqrt{2}}{\beta R}(\partial_2 X^i)^2\right.\right.\cr 
&-&\left.\left.
i\frac{\beta R}{\sqrt{2}2\pi^2\alpha'}\psi^T
\gamma\cdot\partial_1\psi+i2\alpha' B^E\psi^T\partial_1\psi
-\frac{i}{\pi}\psi^T\partial_2\psi
\right]\right\} ~~
\label{zrks}
\eea
where the integration region is the torus of fig.1 .
The boundary conditions for the Bose fields are
\begin{eqnarray}
X^i(\sigma_1+k,\sigma_2)=X^i(\sigma_1,\sigma_2)\cr
X^i(\sigma_1,\sigma_2+r)=X^i(\sigma_1+s,\sigma_2)
\end{eqnarray}
and for the Fermi fields are
\begin{eqnarray}
\psi(\sigma_1+k,\sigma_2)=\psi(\sigma_1,\sigma_2)\cr
\psi(\sigma_1,\sigma_2+r)=(-1)^r\psi(\sigma_1+s,\sigma_2)
\end{eqnarray}

It is possible to change the coordinates in the action (\ref{zrks}) so
that the integration region is the square torus $(\sigma_1, \sigma_2)\in
\left([0,1),[0,1)\right)$.  The appropriate coordinate transformation is
\bea
\sigma_1'&=&\frac{\sigma_1}{k} - \frac{s\sigma_2}{kr}\ ,\cr
\sigma_2'&=& \frac{\sigma_2}{r}
\eea
Then, the action becomes
\be
S=\frac{1}{4\pi\alpha'}
\int_0^1 d\sigma_2\int_{0}^{1}
d\sigma_1\sqrt{g}\left(
g^{\alpha\beta}
\partial_\alpha \vec X\cdot \partial_\beta \vec X-i 2\alpha'\bar\psi^i
\gamma^a e^\alpha_a\partial_\alpha\psi^i \right)
\label{srks}
\ee
where 
\be
g_{\alpha\beta}=\left( \matrix{   1 & \tau_1\cr \tau_1& 
\vert\tau\vert^2\cr}\right)\ 
\ee
with
\be
\tau= \frac{s}{k}+ \tau_0 \frac{r}{k}
\label{taunot}
\ee
where $\tau_0$ is the parameter defined in (\ref{tauzero}) which
was the Teichm\"uller parameter of the basic underlying torus of
DLCQ strings.

$e^\alpha_a$ a zweibein which corresponds to the metric $g_{\alpha\beta}$.
Now, the boundary conditions are
\begin{eqnarray}
X^i(\sigma_1+1,\sigma_2)=X^i(\sigma_1,\sigma_2)\cr
X^i(\sigma_1,\sigma_2+1)=X^i(\sigma_1,\sigma_2)\cr
\psi(\sigma_1+1,\sigma_2)=\psi(\sigma_1,\sigma_2)\cr
\psi(\sigma_1,\sigma_2+1)=(-1)^r\psi(\sigma_1,\sigma_2)
\end{eqnarray}
Note that the boundary condition for the Fermi field still depends on $r$.

We have thus obtained the superstring partition function where the world-sheet 
is a fixed torus of the kind depicted in Figure 1.  This worldsheet is an un-branched
cover of an underlying torus which has Teichm\"uller parameter $\tau_0$ given
in equation (\ref{tauzero}) and coinciding with the basic parameter that we identified
for the superstring.

It is now necessary to show that the sum over all compatible permutations $P$ and $Q$
produce the sum over genus 1 worldsheets which we found for the superstring.  The
relevant arguments are given in ref.\cite{Grignani:1999sp} and we refer the reader there
for the details. An essential piece of the derivation is to show the exponentiation
of the sum over connected parts.  Then, for a connected parts, the sum over the integers
$k,r,s$ in the domain which characterize the permutations produces the sum over discrete Teichm\'uller
parameters which we found for the superstring at genus 1.  In that sum, after replacing the
Euclidean $B^E$ by the Minkowski field $B$, we obtain an expression for the free energy of
the matrix string theory which precisely coincides with  
the superstring partition function which we found in  (\ref{supfe1}).

\section{Hagedorn Temperature}

\subsection{Bosonic String}

In bosonic string theory, the Hagedorn temperature is the temperature 
at which a certain winding modes become massless, indicating a phase 
transition.
We will investigate the Hagedorn transition analyzing the mass shell
condition, looking for the temperature at which a new tachyon, besides 
the one already present at zero temperature, appears \cite{Atick:1988si}.

Let us recall the expression for the world-sheet action in Euclidean space
in a $B^E$-field background.
\begin{equation}
S=
\frac{1}{4\pi\alpha'}\int d^2 {\sigma}\left(
\partial_{\alpha} X^\mu \partial^{\alpha} X_\mu -2\pi i\alpha' B^E_{\mu\nu}
\epsilon^{\alpha\beta}\partial_{\alpha} X^\mu \partial_{\beta} X^\nu\right)
\end{equation}
Having compactified the time dimension to circumference $\beta$ and the 
light-cone direction, one has
\begin{eqnarray}
X^0&=&x^0-(2\alpha ')p^0 i\tau +\left(\frac{m\beta }{\pi }+i\sqrt{2}Rp\right)\sigma +
({\rm oscillators})\cr
X^{1}&=&x^{1}-(2\alpha ')p^{1}i\tau 
+\sqrt{2}Rp\sigma +({\rm oscillators})\cr
X_T^{i}&=&x_T^{i}-2\alpha 'P^i_T i \tau +({\rm oscillators})
\end{eqnarray}

The momenta in the $0$ and $1$ directions are quantized according to 
\begin{eqnarray}
P^0 &\equiv& -\int_0^\pi \frac{\partial \cal L}{\partial i\partial_\tau X^0}
d\sigma \cr
&=&p^0 -B^E\pi \sqrt{2}Rp=\frac{2\pi n}{\beta}\\
P^{1} &\equiv&  -\int_0^\pi \frac{\partial \cal L}{\partial 
i\partial_\tau X^{1}}\cr
&=&p^{1} +B^E \beta m+iB^E \sqrt{2}Rp
=-\frac{\sqrt{2}k}{R} -i\frac{2\pi n}{\beta}
\end{eqnarray}
We then have
\begin{eqnarray}
p^0 &=&\frac{2\pi n}{\beta}
+B^E \pi \sqrt{2}Rp \cr
p^{1}&=&-\frac{\sqrt{2}k}{R} -i\frac{2\pi n}{\beta} 
-B^E \beta m-iB^E \sqrt{2}Rp
\label{alpha0s}
\end{eqnarray}
The mass-shell condition becomes
\begin{eqnarray}
M^2&\equiv&- P^i_T P^i_T = \frac{2}{\alpha '}\left(N+\tilde{N}-2\right)
+\frac{m^2 \beta ^2}
{4\pi ^2 \alpha '^2}+\frac{2k^2}{R^2}+\beta ^2(B^E)^2 m^2 \cr
&+&\frac{2\sqrt{2}B^E \beta }{R}km 
+2\pi i\left[\sqrt{2}\beta Rmp\left(\frac{1}{4\pi ^2 \alpha '^2 }
+(B^E)^2 \right) +\frac{2\sqrt{2}}{R\beta }kn\right] \cr
&+&4\pi iB^E \left(nm+kp\right)
\label{m2}
\end{eqnarray}
The unfamiliar imaginary terms in this equation can be understood
by considering that when $M^2$ is  
exponentiated to obtain the partition function,
and summed over the integers, these terms will provide 
periodic delta functions that constraint the Teichm\"uller parameter.
As a matter of fact there is an alternative way to obtain the mass spectrum. 
One can consider the torus path integral for the partition function.
The sum over path integral topological sectors with the winding mode 
action given by (\ref{swm}) is related by Poisson 
re-summation to the partition function for the shifted spectrum.
By first Poisson re-summing over $n$ and then over $q$ we get
\begin{eqnarray}
&&\sum_{m,n,p,q}\exp\left[-\frac{\beta^2}{4\pi \alpha '
\tau_2}\left(m\bar\tau -n\right)
\left(\tau m-n\right)+2\pi i\frac{\sqrt 2 \beta R}{4\pi \alpha '\tau_2}
\left(nq+|\tau |^2mp \right. \right.\cr
&-&\left.\left. \tau_1(mq+np)\right)-2\pi i\sqrt 2 \beta R\frac{B^E}{2}(mq-np)
\right]\cr
&=&\frac{\sqrt{2}2\pi\alpha' \tau_2}{\beta R}
\sum_{m,k,p,s}\exp\left[-\pi\tau_2\alpha'
\left(\left(\frac{m\beta}{2\pi\alpha'}\right)^2+
\left(\frac{\sqrt{2}k}{R}+\beta B^E m\right)^2\right.\right.\cr
&+&\left.\left.2\pi i\left(\sqrt{2}\beta R m p 
\left((B^E)^2+\frac{1}{4\pi^2\alpha'^2}\right)-\frac{2\sqrt{2}}{R\beta} sk+
2B^E (kp-ms)\right)\right)\right.\cr
&+&\left.
2\pi i\tau_1(kp+ms)\right]
\label{psrsmd}
\end{eqnarray}
This formula clearly reproduces the topological part of 
the mass shell condition (\ref{m2}), when $s\to -n$,
and of the constraint
$L_0 -\tilde{L}_0 =0$, which reads
\begin{equation}
L_0 -\tilde{L}_0 =nm -kp +N-\tilde{N}=0
\end{equation}
with $N$ and $\tilde{N}$ being the left- and right-moving mode 
number operators,
\begin{eqnarray}
N&=&\sum_{n=1}^{\infty}:\alpha_{n}^\mu \cdot \alpha_{-n}^\mu :\cr
\tilde{N}&=&\sum_{n=1}^{\infty}:\tilde{\alpha}_{n}^\mu 
\cdot \tilde{\alpha}_{-n}^\mu :
\end{eqnarray}
From (\ref{psrsmd}) we then get the following form for the mass-shell 
condition
\begin{equation}
M^2=\frac{2}{\alpha '}\left(N+\tilde{N}-2\right)
+\left(\frac{m\beta}{2\pi\alpha'}\right)^2+
\left(\frac{\sqrt{2}k}{R}+m\beta B^E \right)^2
\label{masses}
\end{equation}
and, by summing over $p$ and $s$, the periodic delta-functions
\begin{eqnarray}
&&\delta\left[\pi\alpha'\tau_2\left(\sqrt{2}\beta R m
\left((B^E)^2+\frac{1}{4\pi^2\alpha'^2}\right)+2 B^E k\right)-
\tau_1 k+r\right]\cr
&&\delta\left[2\pi\alpha'\tau_2\left(\frac{\sqrt{2}}{R\beta}k+mB^E \right)+
\tau_1 m-n\right]
\label{ndtp}
\end{eqnarray}
where the sums over $r$ and $n$ have been omitted.
Eq.(\ref{ndtp}) provides
the discretized Teichm\"uller parameter $\tau$. 
As can be easily seen $\tau$ has again the form
$$
\tau=\frac{n(1-2\pi i\alpha'B^E)-ir\nu}{m(1-2\pi i\alpha'B^E)-ik\nu }
$$
So that it coincides with the one defined in (\ref{dtpm})
when $r$ is identified with $s$.

Looking for a state of $N=\tilde{N}=0$, we must set the quantized momenta
to zero, $i.e.$ $n=k=0$, and
to find something ``stringy'',  $m=\pm 1$. Consequently, in Minkowski 
space, $M^2$ becomes
\begin{equation}
M^2=-\frac{4}{\alpha '}+\frac{\beta ^2}{4\pi ^2 \alpha '^2} 
-\beta ^2 B^2
\end{equation}
We see that these modes become tachyonic when
\begin{equation}
\beta^2=\frac{16\pi ^2 \alpha '}{1-(2\pi \alpha 'B)^2}
\label{temp}
\end{equation}
which gives the Hagedorn temperature in the presence of a $B$-field.

For $B \to 0$ we recover the well known 
result \cite{Atick:1988si}
\begin{equation}
T_H =\frac{1}{4\pi \sqrt{\alpha '}}\ .
\end{equation}

\subsection{Superstring}

In this section, we will extend the preceding discussion of the 
Hagedorn transition for bosonic closed string to the case of a
supersymmetric type II superstring in a constant $B$-field.

The description of supersymmetric string theories in the 
$NSR$ formalism requires the sum over spin structures in 
addition to summing over windings.
For type II superstrings we must consider the contribution of both 
right-moving and left-moving $NSR$ fermions.
At genus one 
there are four spin structures for the right- and left-moving fermions.
These spin structures are $(+,+)$, $(+,-)$, $(-,+)$, $(-,-)$, where 
the first sign refers to the $\sigma_1$ direction and the second to 
the $\sigma_2$ direction.
Thinking of $\sigma_1$ and $\sigma_2$ as the world-sheet space and 
time direction, one has that space-time bosons ($R$ sector) correspond to 
$(+,+)$ and $(+,-)$ and space-time fermions ($NS$ sector) 
correspond to $(-,+)$ and $(-,-)$.

In \cite{Atick:1988si}, is explained that the effective 
string theory that governs the high-temperature
limit of type II superstrings is a theory in which one sums over all 
spin structures $L$, imposing that the right-moving spin structure
is the same as the left-moving one.
This is a ghost-free and modular invariant but tachyonic ten-dimensional 
string theory.

In order to study the Hagedorn temperature for type II superstring, we must
look for a state which is in the $NS-NS$ sector, because only in this 
case the tachyon appears.

As shown in the superstring section, the $B$-dependent contribution 
comes from the bosonic part of the action. Thus
the formula for the mass-shell condition 
is similar to that obtained for the bosonic string, with $N$ and 
$\tilde{N}$ being respectively the right- and left-moving 
number operators of bosons and fermions, and with the appropriate 
value for the normal ordering constant (which is $1/2$ for 
both right and left sectors)
\begin{equation}
M^2=\frac{2}{\alpha '}\left(N+\tilde{N}-1\right)
+\left(\frac{m\beta}{2\pi\alpha'}\right)^2+
\left(\frac{\sqrt{2}k}{R}+i\beta B m\right)^2
\end{equation}
We are looking for a state with 
$N=\tilde{N}=0$, so we must set the quantized momenta to zero,
and, to find something ``stringy'', $m=\pm 1$.
Except for the presence of the $B$-field in the mass-shell condition,
all the discussion proceeds as in \cite{Atick:1988si}.
In Minkowski space the mass-shell condition becomes,
\begin{equation}
M^2=-\frac{2}{\alpha '}+\frac{\beta ^2}{4\pi ^2 \alpha '^2} 
-\beta ^2 B^2
\end{equation}
These modes become tachyonic when
\begin{equation}
\beta^2=\frac{8\pi ^2 \alpha '}{1-(2\pi \alpha 'B)^2}
\label{temps}
\end{equation}
which gives the Hagedorn temperature in the presence of a $B$-field.

For $B \to 0$ we recover the well known 
result \cite{Atick:1988si}
\begin{equation}
T_H =\frac{1}{2\pi \sqrt{2\alpha '}}\ .
\end{equation}

It is possible to reproduce the result given in 
(\ref{temps}) by studying where the free
energy in the Green-Schwarz 
formulation of the superstring diverges.

\section{Discussion}

What we have examined in this Paper is an interesting coincidence between the
set of all eigenvalues of simultaneously diagonalizable matrices which live on
a torus and the set of Riemann surfaces that appear in the path integral for 
the string partition function.  In both cases they are the sets of branched
covers of a basic torus.  We have found the further fact that the basic torus
is the same in the case of matrix theory and the type II superstring.  We also
find that the one-loop, or genus one, contribution in both cases are identical
in every detail.

It would be interesting to extend this correspondence to higher genera.  We have
done this partially by showing that the set of branched covers of the torus
is what appears in the sum over Riemann surfaces in the string partition function.
In this case, the higher genus covers can be obtained by gluing tori together along
branch cuts. Analogous to this, 
it should be possible to show that a higher genus 
contribution to the string partition
function can be obtained by taking correlators of the DVV vertex operators
in the free matrix string model.  This would potentially establish the 
appearance of string perturbation theory within the matrix model to all orders
in the genus expansion.

The effect of a $B$-field on the Hagedorn temperature that we have found is interesting.
$T_H$ depends on $B$ and is lowered by increasing $B$, but it does not depend on the
light-cone compactification radius $R$.  This means that, for non-zero $B$, the DLCQ
procedure has an infinite $R$ limit which remembers the compactification and still
feels the effect of $B$.  This puzzling behavior needs further investigation.  It
leaves open the possibility that the DLCQ and at finite temperature is inconsistent
in the sense that $B$ doesn't decouple properly as the physically interesting limit
$R\to\infty$ is taken.

\section*{Acknowledgements}
We thank Mark Van Raamsdonk and Moshe Rozali for many useful discussions.
G. Semenoff and M. Orselli acknowledge the hospitality of the 
Dipartimento di Fisica, Universit\`a di Perugia and G. Grignani the Pacific 
Institute for Mathematical Sciences where parts of this work were 
completed.

\end{document}